\documentclass[pdfLatex,twocolumn,epjc3]{svjour3}          
\RequirePackage[T1]{fontenc}
\smartqed  
\RequirePackage{graphicx}
\RequirePackage{mathptmx}
\RequirePackage{flushend}
\RequirePackage[numbers,sort&compress]{natbib}
\RequirePackage[colorlinks,citecolor=blue,urlcolor=blue,linkcolor=blue]{hyperref}
\usepackage{amsmath}
\usepackage{multicol}
\usepackage[english]{babel}
\usepackage{amsmath}
\usepackage{amssymb}
\usepackage{babel,blindtext}
\usepackage{float}
\usepackage{amsfonts}
\usepackage{graphicx}
\usepackage[all]{xy}
\usepackage{tabularx}
\usepackage{verbatim}
\usepackage{graphicx}
\usepackage{dcolumn}
\usepackage{bm}
\usepackage{subcaption}
\usepackage{hyperref}
\usepackage[mathlines]{lineno}
\usepackage{dcolumn}
\usepackage{amsmath}
\usepackage{amsfonts}
\usepackage{array}
\usepackage{booktabs}
\usepackage{tabu}
\usepackage{dcolumn}
\usepackage{amsmath}
\usepackage{amsfonts}
\usepackage{amssymb}
\usepackage{graphicx}
\usepackage{mathtext}
\usepackage{graphicx}
\usepackage{dcolumn}
\usepackage{bm}
\usepackage{mathtools}
\usepackage{hyperref}
\usepackage{natbib}
\newcommand{\be}{\begin{equation}}
\newcommand{\ee}{\end{equation}}
\newcommand{\ber}{\begin{eqnarray}}
\newcommand{\eer}{\end{eqnarray}}
\newcommand{\bern}{\begin{eqnarray*}}
\newcommand{\eern}{\end{eqnarray*}}
\newcommand{\beast}{\begin{equation*}}
\newcommand{\eeast}{\end{equation*}}

\begin{document}
\title{Computation of Thermodynamic and Hydrodynamic Properties of the Viscous 
      Atmospheric Motion on the Rotating Earth in 2D Using Naiver-Stokes Dynamics}

\author{Admasu Abawari \thanksref{addr1, addr2, e1}
        \and
        Yitagesu Elfaged \thanksref{addr3, addr1,e2}
}
\thankstext{e1}{admasu.abawari@aau.edu.et}
\thankstext{e2}{yitagessu.elfaged@aau.edu.et}
\institute{Department of Physics, Addis Ababa University, P.O. Box 1176, Addis Ababa, Ethiopia \label{addr1}\and Department of Physics, Mizan-Tepi University,P.O. Box 260, Mizan Teferi, Ethiopia \label{addr2}
          \and
         Department of Physics,  Addis Ababa University, P.O. Box 1176, Addis Ababa, Ethiopia \label{addr3}
          }
\date{\today}

\maketitle

\begin{abstract}
In this article, we model Earth's lower small-scale eddies motion in the atmosphere as a compressible neutral
fluid flow on a rotating sphere. To justify the model, we carried out a numerical computation of the thermodynamic
and hydrodynamic properties of the viscous atmospheric motion in two dimensions using Naiver-Stokes dynamics,
 conservation of atmospheric energy, and continuity equation. The dynamics of the atmosphere, governed by partial differential equation
without any approximation ,and without considering latitude-dependent acceleration due to gravity. The numerical
solution for those governed equations was solved by applying the finite difference method with applying some sort
of horizontal air mass density as a perturbation to the atmosphere at a longitude of $5\Delta\lambda$ . Based on this initial
boundary condition with taking temperature-dependent transport coefficient in to account, we obtain the propagation for
each atmospheric parameter and presented in graphically as a function of geometrically position and time. All of the parameters oscillating with respect to time and satisfy the characteristics of atmospheric wave. 
 Finally, the effect of the Coriolis force on resultant velocity were also discussed by plotting
contour lines for the resultant velocity for different magnitude of Coriolis force, then we also obtain an interesting wave
phenomena for the respective rotation of the Coriolis force.

\end{abstract}
  
~~~~Keywords: Naiver-Stokes Equations; Finite difference method; Viscous atmospheric motion; Viscous dissipation; convective motion.
\section{Introduction}\label{introduction1}
As stated in \cite{andrews2010introduction},
it is very difficult to define the exact size, mass, and weight of the Earths atmosphere,
but in many respects, our atmosphere is a very thin layer, compared to Earth's radius. It is a critical system for life on our planet and together with the oceans, the atmosphere shapes Earth's climate and weather patterns and makes some regions more habitable than others. The atmosphere
consists of a mixture of ideal gases: although molecular nitrogen and molecular oxygen
predominate by volume.
Like other planetary atmospheres, the earth's atmosphere figures centrally in transfers of energy between the sun and the planet's surface and from one region of the globe to another; these transfers maintain thermal equilibrium and determine the planet's climate \cite{curry1998thermodynamics}.\\
\\
As cited in reference \cite{denur2016condensation}~, Atmospheric thermodynamics is the study of heat-to-work transformations(and their reverse) that take place in the earth's atmosphere and manifest as the weather of climate.
Hence it is involved in every atmospheric process, from the large-scale general circulation to the local transfer of radiative, sensible and latent heat between the surface and the atmosphere and the microphysical processes producing clouds \cite{cairo2011atmospheric}.   
 As we stated above, the earth's atmosphere is the  fluid system envelope surrounding the planet, and the atmosphere is capable of supporting a wide spectrum of
motions, ranging from turbulent eddies of a few meters to circulations having dimensions
of the earth itself. Its motion is strongly influenced by different factors such as the
effect of the rotation of the Earth, gravity of the earth, air pressure force and the viscosity of the fluid \cite{salby1996fundamentals}.
\\
\\
As suggested in reference~ \cite{vallis2016geophysical}, the atmosphere is governed by the laws of
continuum mechanics and these can be derived from the laws of mechanics and thermodynamics governing a discrete fluid body by generalizing those laws to a continuum of
such systems. In order to simulate these fluid flow and other related physical phenomena,
it is necessary to describe the associated physics in mathematical form.
\\
\\
Nearly all the physical phenomena of interest to us in this finding are governed by principles of conservation and are expressed in terms of partial differential equations. Hence, Cloaude-Louis ~Navier (1823) derived the equations of motion for a viscous fluid
from molecular considerations, and George Stokes (1845) also derived the equations of
motion for a viscous fluid in a slightly different form and the basic equations that govern fluid flow are now generally known as the \textbf{Naiver-Stokes equations} which arise from
applying Isaac Newton's second law to fluid motion~ \cite{childs2010rotating}. According to \cite{wallace2006atmospheric}, the dynamics
that described by the Naiver-Stokes equations (NSE) of fluid dynamics, for a variable
the density of incompressible ocean and compressible atmosphere expressing conservation of
mass, momentum, and energy. 
All the atmospheric gases are constituents of air characteristics with their pressure, density, and temperature  and these parameters vary
with altitude, latitude, longitude, and related to each other by the
equation of state \cite{jacobson2005fundamentals} and  these field variables are related to one another by the equation of state for
an ideal gas.\\
\\
Life is too short to solve every complex problem in detail, and the atmospheric and oceanic
sciences abound with such a complex problem. To solve real-world problems we need to
add water vapor as well as the equations of radiative transfer on Thermodynamical equation. All these make a complex system, and to make progress, we need to simplify where
possible and eliminate unimportant effects from the model. Thus, during the last decades,
direct numerical simulations (DNS) have been recognized as a powerful and reliable tool
for studying the basic physics of turbulence, and numerous findings showed that the results obtained by DNS are in excellent agreement with experimental findings \cite{nikitin2006finite}.  
In reference, \cite{burger2010numerical},
one of the challenging equations in atmosphere phenomena is the Naiver-Stokes equations
which can not be solved exactly. So, approximations and simplifying assumptions are
commonly made to allow the equations to be solved approximately. \\
\\
Recently, high speed
computers have been used to solve such equations by replacing them with a set of algebraic
equations using a variety of numerical techniques like finite difference method (FDM) \cite{reis2015compact,
tsega2018finite},~the finite element method (FEM) \cite{kolmogorov2014finite} the finite volume method (FVM)\cite{nath2016numerical}, meshfree methods and boundary elements method \cite{sedaghatjoo2018numerical}.
Many authors in their finding  considered  the coefficient  of viscosity as well as  the thermal conductivity of the fluid   as a constant parameter, and  on many other studies fluid as incompressible for numerical computation of Navier-Stokes dynamics,and there is still no findings that have 
been done considering on this area (apply the temperature dependent viscosity and thermal conductivity and their influence on the accuracy of the final
solution ) for solving an atmosphere model, and we numerical compare on the two solutions through finding the relative error between on two mechanism.\\
\\
The strategy intended to achieve our goal that is to compute the thermodynamic and hydrodynamic properties of the viscous atmospheric motion on the rotating Earth frame. The assumptions used in this analysis are the flow is simple a compressible   neutral fluid,  with considering temperature dependent transport coefficients of the fluid flow over  small-scale motion. In order to get the distributions of the resultant velocity, pressure, density, and temperature of the fluid on our atmosphere, the governing equations have been derived based on the above-mentioned assumptions and given as follows in Section\ref{atmospheric} and \ref{nse}, and the numerical results obtained are presented graphically and
discussed. in section \ref{dis}.
\section{The Laws of Thermodynamics of the Atmosphere }\label{atmospheric} 
In this context, the first law of the thermodynamics which when applied to the moving fluid elements is consists of the net transfer of heat by fluid flow,  net heat transfer by conduction,
rate of internal heat generation, 
the rate of work transfer from the control volume to its environment. Hence, the first law of thermodynamics equation containing the viscous dissipation term $\varPhi$ is given in equation \eqref{eq0}
\begin{equation}\label{eq0}
 \rho\frac{D e}{Dt}=\rho\dot{q}+k\nabla^{2}T+\nabla k\nabla T-p\nabla\bullet\vec{u}+\varPhi
\end{equation}
The way we have written \eqref{eq0} is intended to give a clear picture of the balances of
work and energy in the flow field. Thus, the heat added by conduction, radiation and chemical
reaction (the right-hand-side terms in the first line of \eqref{eq0} is employed directly to increase
the total internal energy. Viscous dissipation as well as compression work, written for clarity
in separate lines in \eqref{eq0}, are the mechanisms transforming mechanical energy into
internal energy, with only the latter being reversible\cite{monteiro2007dynamic}.\\
\\
The natural coordinates in which to express our equations, when they are applied to the
 Earth, are spherical coordinates $(r, \phi, \lambda)$, where $r$ is the distance from the center of the
 Earth,~ $\phi$ is latitude and $\lambda$ is longitude. The detailed derivation of the atmospheric equations of motion \eqref{eq0} in Cartesian coordinates was given in \cite{andrews2010introduction},
and in spherical coordinate for an compressible fluid it  becomes expressed as
\begin{equation}\label{eq1}
\begin{split}
 &\rho c_{ v }\left[\frac{\partial T}{\partial t}+\frac{u}{r\cos\phi}\frac{\partial T}{\partial \lambda}+\frac{v}{r}\frac{\partial T}{\partial \phi}+w\frac{\partial T}{\partial z}\right]=\\& k\left[
 \frac{1}{r^{2}\cos^{2}\phi}\frac{\partial^{2} T}{\partial \lambda^{2}}-\frac{\tan\phi}{r^{2}}\frac{\partial T}{\partial \phi}+\frac{2}{r}\frac{\partial T}{\partial z}+\frac{1}{r^{2}}\frac{\partial^{2} T}{\partial \phi^{2}}+
 \frac{\partial^{2} T}{\partial z^{2}}\right]\\&+\left(\frac{1}{r\cos\phi}\right)^{2}\frac{\partial k}{\partial \lambda}\frac{\partial T}{\partial \lambda}+\frac{1}{r^{2}}\frac{\partial k}{\partial \phi}\frac{\partial T}{\partial \phi}+\frac{\partial k}{\partial r}\frac{\partial T}{\partial r}\\&-p\left(\frac{1}{r\cos\phi}\frac{\partial u}{\partial \lambda}+\frac{1}{r}\frac{\partial v}{\partial \phi}+\frac{2w}{r}-\frac{v\tan\phi}{r}-\frac{u}{r}+\frac{\partial w}{\partial r}\right)\\&\rho\dot{q}+\varPhi_{I}
 \end{split}
\end{equation}
Here, the coefficient of the thermal conductivity of the air ,and it's  viscosity $\mu$ 
depends on the temperature based on experimental result and they are approximately by using Sutherland's law\cite{smits2006turbulent}
\begin{equation}\label{e3}
 k_{T}=k_{o}\left(\frac{T}{T_{o}}\right)^{\frac{3}{2}}\frac{T_{o}+S_{k}}{T+S_{k}}
\end{equation}
\begin{equation}\label{e4}
 \mu_{T}=\mu_{o}\left(\frac{T}{T_{o}}\right)^{\frac{3}{2}}\frac{T_{o}+S_{\mu}}{T+S_{\mu}}
\end{equation}
Where, $S_{k},\& S_{\mu}$ are  an effective temperature for cofficient of thermal conductivity and viscosity of the air ,respectively and their  value as mentioned in Sutherland law.  And  $ \mu ~,~ \mu_{o} $ are dynamic viscosity at input temperature T and  at reference temperature $T_{o}$ ,respectively.
In the atmospheric range of temperatures, the two specific heat capacity   are given by $c_{p} = c_{pd}\left(1+0.87q_{v}\right)$ and  $c_{v} = c_{pv}\left(1+0.97q_{v}\right)$  where,$q_{v}$ is the mass ratio of the vapor to the moist air ($q_{v}=\frac{m_{v}}{m_{v}+m_{d}}$).  The dissipation term $\varPhi_{I}$ in spherical coordinate system for compressible fluid flow is given as 
\begin{equation}\label{eq2}
\begin{split}
 &\varPhi_{I}=\left(2\mu-\upsilon\right)\left(\frac{1}{r\cos\phi}\frac{\partial u}{\partial \lambda}+\frac{1}{r}\frac{\partial v}{\partial \phi}+\frac{2w}{r}-\frac{v\tan\phi}{r}\right)\\&+\left(2\mu-\upsilon\right)\left(-\frac{u}{r}+\frac{\partial w}{\partial r}\right)+\frac{\mu}{r^{2}}\left(\frac{\partial u}{\partial\phi}+\frac{1}{\cos\phi}\frac{\partial v}{\partial \lambda}\right)^{2}\\&-\frac{4\mu}{r\cos\phi}\frac{\partial u}{\partial\lambda}\frac{\partial v}{\partial\phi}
 \end{split}
\end{equation}
\section{The Naiver-Stokes Equations for the moist air in Rotational Coordinate System}\label{nse}   
In this section, we introduce the basic fluid dynamical laws that govern our atmospheric flows. When we assume that the blob is instantaneous of cuboidal shape,
with sides $\delta x, \delta y$ and $\delta z$ which is fixed in space and different type of forces exerted on this  blob, then Newton's second could be expressed as 
\begin{equation}\label{eq3}
	m\frac{d\vec{U}}{dt}=\sum \vec{F}
\end{equation} 
Here, the vector $\vec{F}$ is the sum of all the relevant forces exerted on the neutral fluid elements that include the pressure, viscous, Coriolis, and effective gravitational forces. Hence, the equations governing small-scale atmospheric motion would be also derived
from a Lagrangian perspective by considering the rotation of the Earth which adopted from ref.\cite{andrews2010introduction}, and in vector form:  
\begin{equation}\label{eq4}
 \rho\frac{D\vec{U}}{Dt}=-\frac{{\nabla} p}{\rho}-2{\Omega}\times U+\rho\vec{g}+\nabla\bullet\Bbbk
 \end{equation}   
In order to formulate the last term of \eqref{eq4}, we use the following analytic
expression for the stress tensor$     ~\Bbbk~$\cite{zdunkowski2003dynamics}  :
\begin{equation}
 \Bbbk=\mu\left(\nabla\vec{v}+\vec{v}^{\curvearrowleft}\nabla\right)-\lambda\nabla\bullet\vec{v}{E}
\end{equation}
We have  used Lam\`{e}'s coefficients of viscosity $\lambda$ and $\mu$,
which will be treated as temperature-dependent.
\begin{equation}\label{eq25}
\nabla\bullet\Bbbk=\mu\nabla^{2}\vec{v}+\left(\nabla-\lambda\right)\nabla\left(\nabla\bullet\vec{v}\right)+\nabla\mu\bullet\left(\nabla\vec{v}+\vec{v}^{\curvearrowleft}\nabla\right) -\left(\nabla\bullet\vec{v}\right)\nabla\lambda
\end{equation}
In Cartesian coordinator, the viscous forces along x and y components from equation\eqref{eq25} can be expressed as following in equations \eqref{eq26},and \eqref{eq27}, respectively.
\begin{equation}\label{eq26}
 \nabla\bullet\Bbbk_{x}=\mu\nabla^{2}u+\left(\mu-\lambda\right)\frac{\partial}{\partial x}\left(\nabla\bullet\vec{v}\right)+2\left(\frac{\partial \mu}{\partial x}\frac{\partial\vec{u}}{\partial x}\right)+2\frac{\partial\mu}{\partial y}\left(\frac{\partial u}{\partial y}+\frac{\partial v}{\partial x}\right)-\frac{\partial\lambda}{\partial x}\left(\nabla\bullet\vec{v}\right)
\end{equation}
\begin{equation}\label{eq27}
 \nabla\bullet\Bbbk_{y}=\mu\nabla^{2}v+\left(\mu-\lambda\right)\frac{\partial}{\partial y}\left(\nabla\bullet\vec{v}\right)+2\left(\frac{\partial \mu}{\partial y}\frac{\partial\vec{v}}{\partial y}\right)+2\frac{\partial\mu}{\partial x}\left(\frac{\partial u}{\partial y}+\frac{\partial v}{\partial x}\right)-\frac{\partial\lambda}{\partial y}\left(\nabla\bullet\vec{v}\right)
\end{equation}
In spherical coordinates
with the components of the velocity vector given by
$\left(\vec{u}=u ,v \right)$
 , the Naiver-Stokes equations are given by
 \begin{itemize}
 \item  \textbf{$\lambda$- component of the momentum equation}
 \begin{equation}\label{eq5}
 \begin{split}
  &\frac{\partial u}{\partial t}+\left(\frac{u}{r\cos\phi}\frac{\partial u}{\partial\lambda}+\frac{v}{r}\frac{\partial u}{\partial\phi}\right)+\frac{uw}{r}-\frac{uv}{r}\tan\phi+lw-fv\\&+\frac{1}{r\rho \cos\phi}\frac{\partial p}{\partial \lambda}=\\&\mu\left(\frac{4}{3r^{2}\cos^{2}\phi}\frac{\partial^{2} u}{\partial \lambda^{2}}-\frac{\tan\phi}{r^{2}}\frac{\partial u}{\partial \phi}+\frac{1}{r^{2}}\frac {\partial^{2} u}{\partial \phi^{2}}\right)\\&+\frac{\left(\mu-\lambda\right)}{r\cos\phi}\frac{\partial}{\partial \lambda}\left(\nabla\bullet\vec{v}\right)-\frac{1}{r\cos\phi}\frac{\partial \lambda}{\partial \lambda}\left(\nabla\bullet\vec{v}\right)+\frac{2}{r^{2}\cos^{2}\phi}\frac{\partial\mu}{\partial\lambda}\left(\frac{\partial u}{\partial \lambda}\right)+\\&\frac{2}{r^{2}}\frac{\partial \mu}{\partial \phi}\left(\frac{\partial u}{\partial \phi}+\frac{1}{\cos\phi}\frac{\partial v}{\partial \lambda}\right)
  \end{split}
 \end{equation}
\item  \textbf{$\phi$- component of the momentum equation}
 \begin{equation}\label{eq6}
 \begin{split}
  &\frac{\partial v}{\partial t}+\left(\frac{u}{r\cos\phi}\frac{\partial v}{\partial\lambda}+\frac{v}{r}\frac{\partial v}{\partial\phi}\right)+\frac{vw}{r}-\frac{u^{2}}{r}\tan\phi+\\&fu+\frac{1}{r\rho }\frac{\partial p}{\partial \phi}=\mu\left(\frac{4}{3r^{2}}\frac{\partial^{2} v}{\partial \phi^{2}}+\frac{1}{r^{2}\cos^{2}\phi}\frac{\partial^{2}v}{\partial \lambda^{2}}\right)\\&+\frac{\left(\mu-\lambda\right)}{r}\frac{\partial}{\partial \phi}\left(\nabla\bullet\vec{v}\right)-\frac{1}{r}\frac{\partial \lambda}{\partial \phi}\left(\nabla\bullet\vec{v}\right)+\frac{2}{r^{2}}\frac{\partial\mu}{\partial\phi}\left(\frac{\partial u}{\partial \phi}\right)+\\&\frac{2}{r^{2}\cos\phi}\frac{\partial \mu}{\partial \phi}\left(\frac{\partial u}{\partial \phi}+\frac{1}{\cos\phi}\frac{\partial v}{\partial \lambda}\right)
  \end{split}
 \end{equation}
\end{itemize}
Here,the latitudinal dependence of the Coriolis parameter
is accounted for this work are $ f = 2\Omega \sin \phi $ , $ l = 2\Omega \cos \phi$ , respectively and $\rho$ 
is the density of the fluid, $p$ pressure, $\vec{g}$ the effective gravitational acceleration, and the right side terms of equation\eqref{eq5}-\eqref{eq6} comes from  the normal and
shear stresses due to friction.\\
\\
\section{Application of the Dynamics}
As cited in ref~\cite{saks2011cauchy},~the above  both atmospheric energy ,and Naiver-Stokes equations are useful because they describe the physics of many phenomena of scientific and engineering interest. They may be used to model the weather, ocean currents, water flow in a pipe, and airflow around a wing. The Naiver-Stokes equations, in their full forms, help with the design of aircraft and cars, the study of blood flow, the design of power stations, the analysis of pollution, and many other things. Coupled with Maxwell's equations, they can be used to model and study magneto-hydrodynamics.\\
\\
The above description of the fluid system is not complete until we also provide a relation between
density and pressure for the heterogeneous nature of the Earth's lower atmospheric layer. For moist neutral air in the atmosphere behaves approximately as an ideal gas, and its density depends on the specific humidity and virtual temperature of the air, and so we write:
\begin{equation}\label{10}
 p=\rho R_{d}T_{o}\left(1+0.608q_{v}\right)
\end{equation}
We  recall  that  global  existence  of solutions  of  the  atmospheric energy and the Naiver-Stokes  equations  is known to hold in every space and time dimension
\cite{georgiev2018existence}. However, theoretical understanding of the solutions to these equations in many finding is done by taking some standard techniques to estimate wind resources and they did not considered  tiny errors at the smallest scales will ultimately have huge effects on the overall solution.
\section{Numerical Simulation of the Primitive Equations}\label{scale}
In this section, we propose finite-difference-central difference schemes, Runga Kutta fourth-order schemes, and 2D unstaggered grid methods were implemented for proving of existence of global solutions for the above governed equations. 
In all numerical solutions, the continuous partial differential equation  is replaced with a discrete approximation, and
the discretization of those equations are
carried out with respect to dimensional coordinates $\lambda$,and $\phi$ to convey the equations in finite difference form by approximating the functions and their derivatives in terms of central differences with applying the Courant-Friedrichs-Lewy stability criterion be obeyed\cite{zdunkowski2003dynamics}, and the spherical form of this criterion we have used for numerical solve our model is given
by ~$\frac{\partial t}{r^{2}\cos^{2}\phi\partial \lambda^{2}},\frac{\partial t}{r^{2}\partial \phi^{2}}\leq\frac{1}{4}$. \\
\\
As indicated in last section of the introduction, we  considered  the small scale neutral fluid motion of lower atmosphere as a compressible. Our flow is a shallow box of dimensions $ lx=45^{o}$ in the longitude direction,and $ ly=45^{o}$ in the latitude direction  of the earth. The grid space is $\Delta \lambda=1^{-1}$ in the longitude direction,and $\Delta \phi=1^{-1}$ in the latitude direction  of  the Earth, with these parameters,  the angular speed of the rotation of Earth  to be $\Omega=7.30*10^{-5} s^{-1}$. We apply some sort of horizontal air mass density($\rho=10kg/m^{3}$) as a perturbation to the atmosphere at a longitude of $5\Delta\lambda$,and  some useful physical constants were taken from Appendix A of reference\cite{andrews2010introduction}. The computations have been performed using $\Delta t = 1$ to ensure the stability with the time step $n_{t}=50000$, then the over all characteristics of the atmospheric variables would be brief discussed in the next section \ref{dis}.
\newpage
\section{Results and Discussion}\label{dis}
\begin{figure*}[ht!]
    \begin{subfigure}[t]{0.5\textwidth}
        \centering
        \includegraphics[height=2.40in]{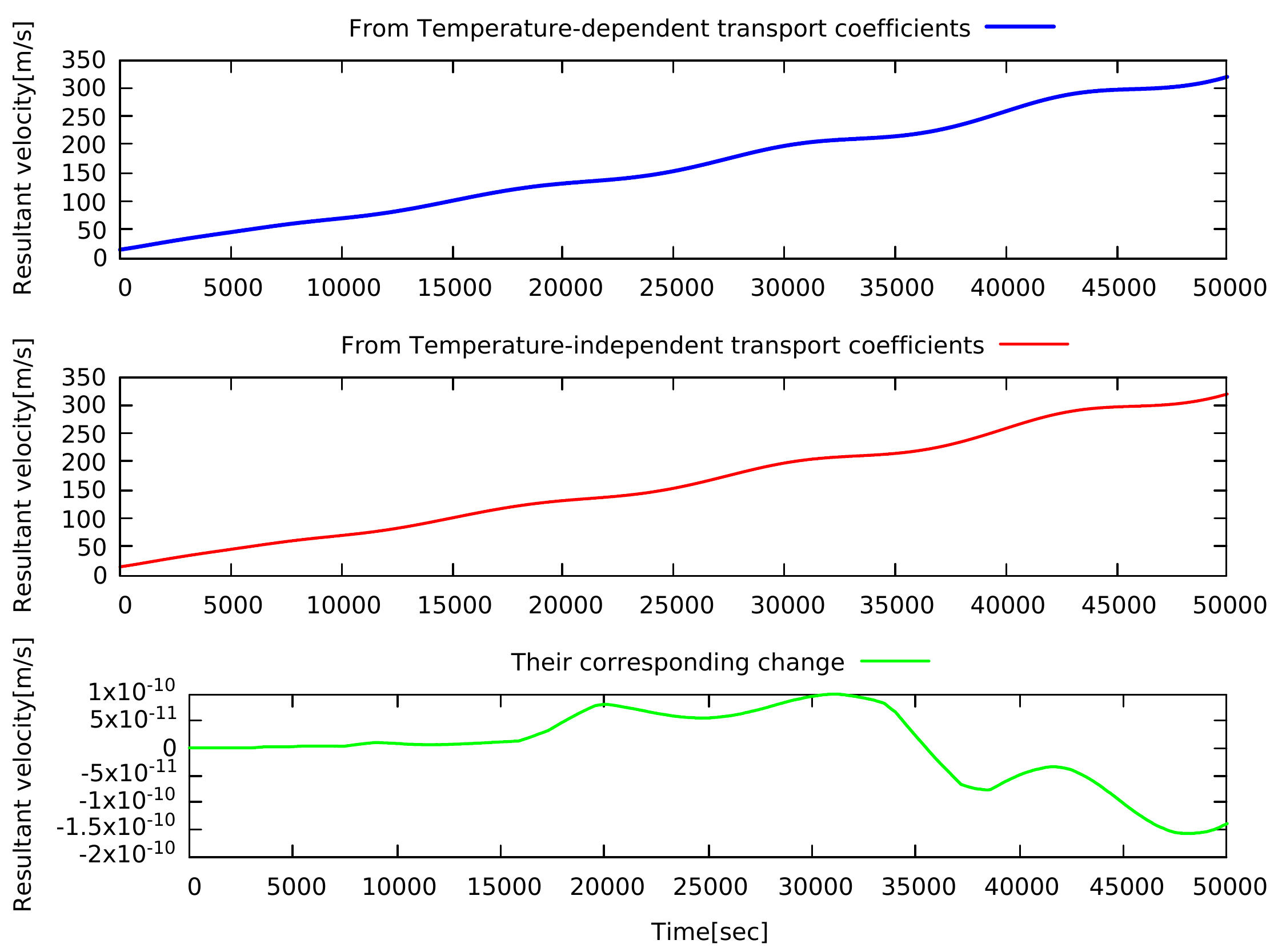}
        \caption{\label{fig1a} This figure shows how $\vec{V}$ changes during iteration steps. }
    \end{subfigure}
    \begin{subfigure}[t]{0.5\textwidth}
        \centering
        \includegraphics[height=2.40in]{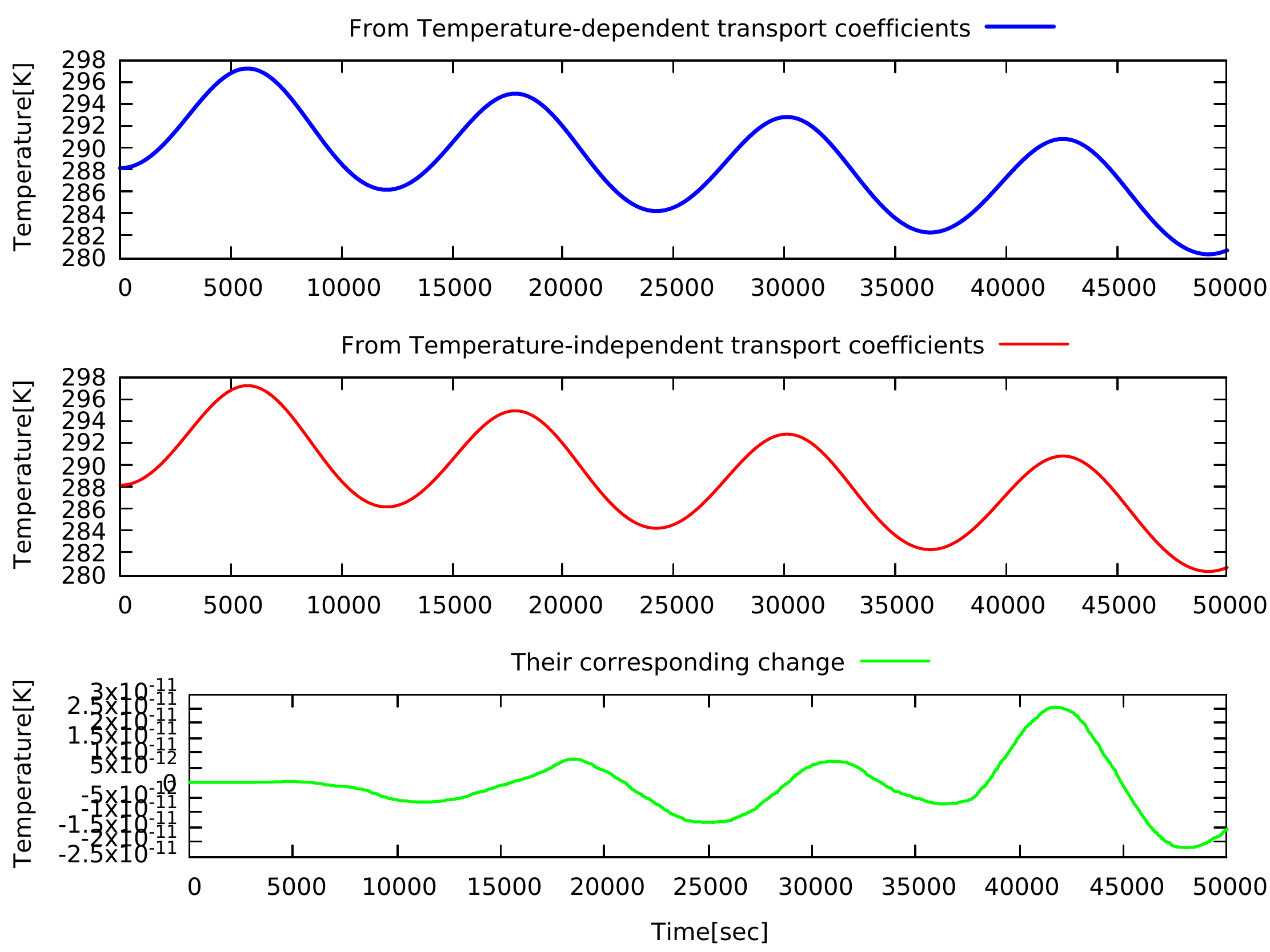}
        \caption{\label{fig1b} This figure shows how temperature changes during iteration steps.}
    \end{subfigure}
    \caption{\label{fig1} 
    Propagation of resultant velocity and the temperature of moist air with respect to time. From top to bottom:
resultant velocity(Fig.$\ref{fig1a}$) and temperature (Fig.$\ref{fig1b}$)}
    \end{figure*}

\begin{figure*}[ht!]
     \begin{subfigure}[t]{0.5\textwidth}
        \centering
        \includegraphics[height=2.40in]{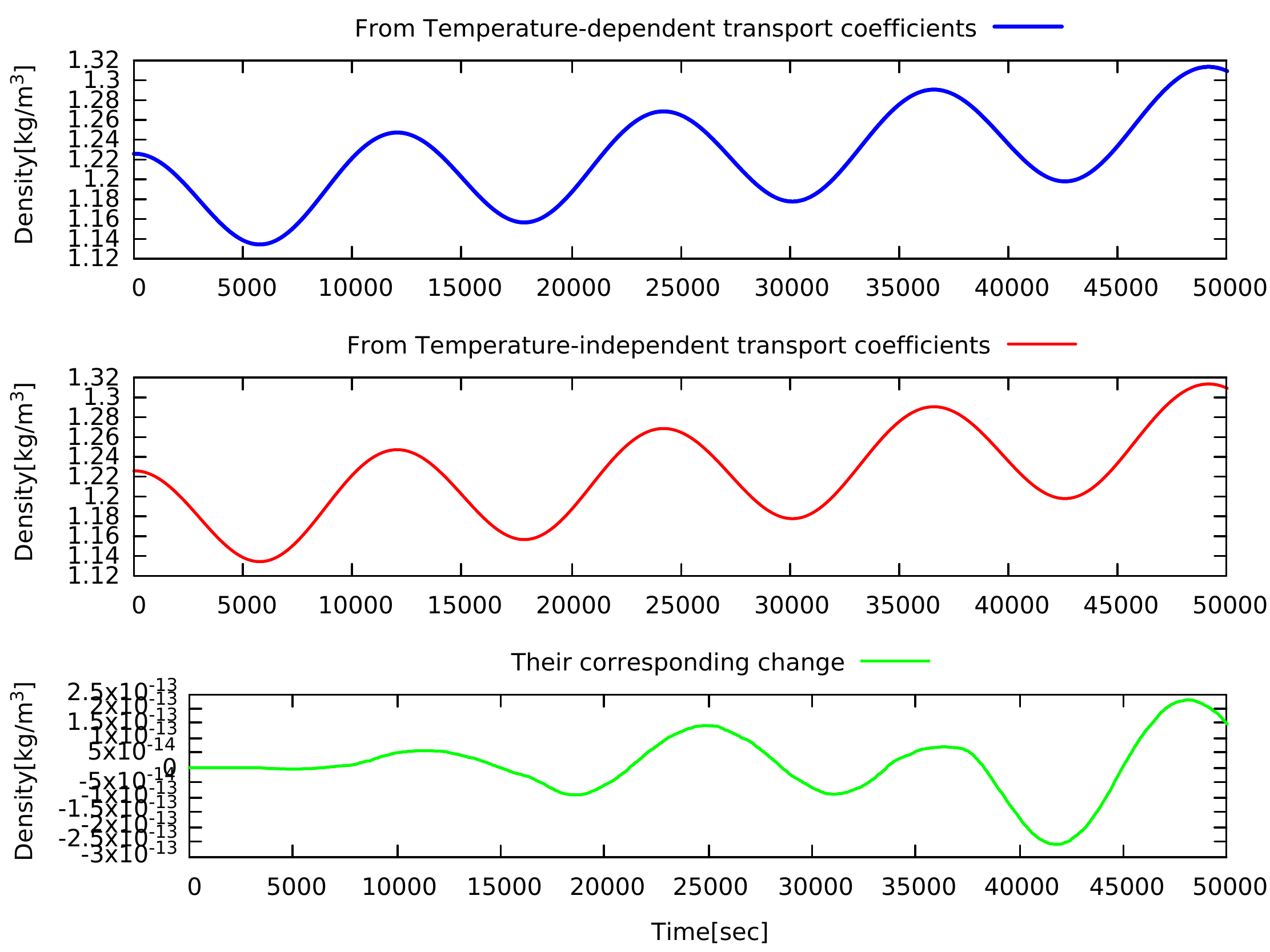}
        \caption{\label{fig2a} This figure shows how density changes during iteration steps. }
    \end{subfigure}
    \begin{subfigure}[t]{0.5\textwidth}
        \centering
        \includegraphics[height=2.40in]{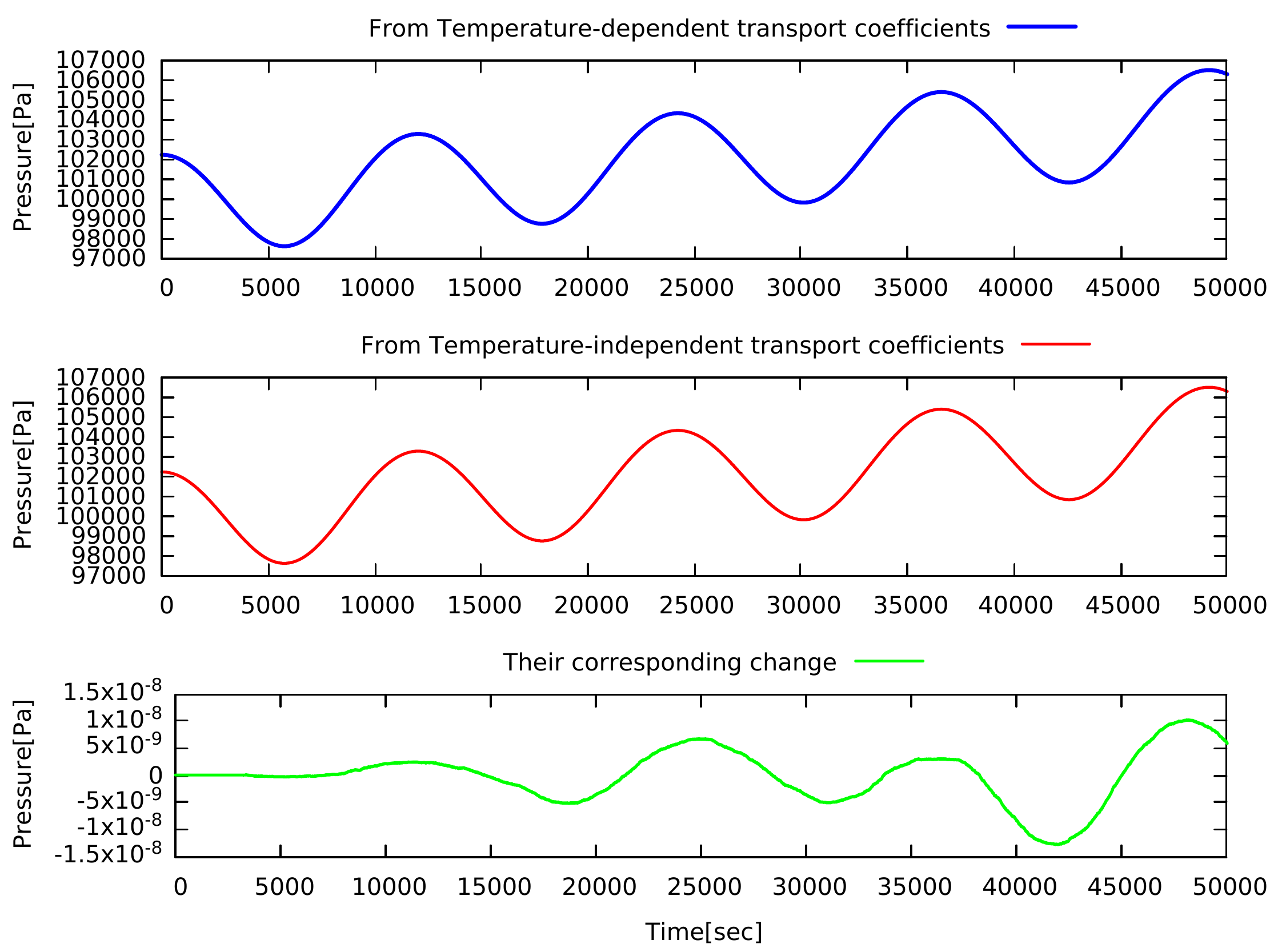}
        \caption{\label{fig2b} This figure shows how pressure changes during iteration steps.}
    \end{subfigure}
    \caption{\label{fig2} 
    Propagation of density and pressure of moist air with respect to time. From top to bottom: density(Fig.$\ref{fig2a}$)
and pressure (Fig.$\ref{fig2b}$)}   
\end{figure*}

\begin{figure*}[ht!]
     \begin{subfigure}[t]{0.5\textwidth}
        \centering
        \includegraphics[height=2.40in]{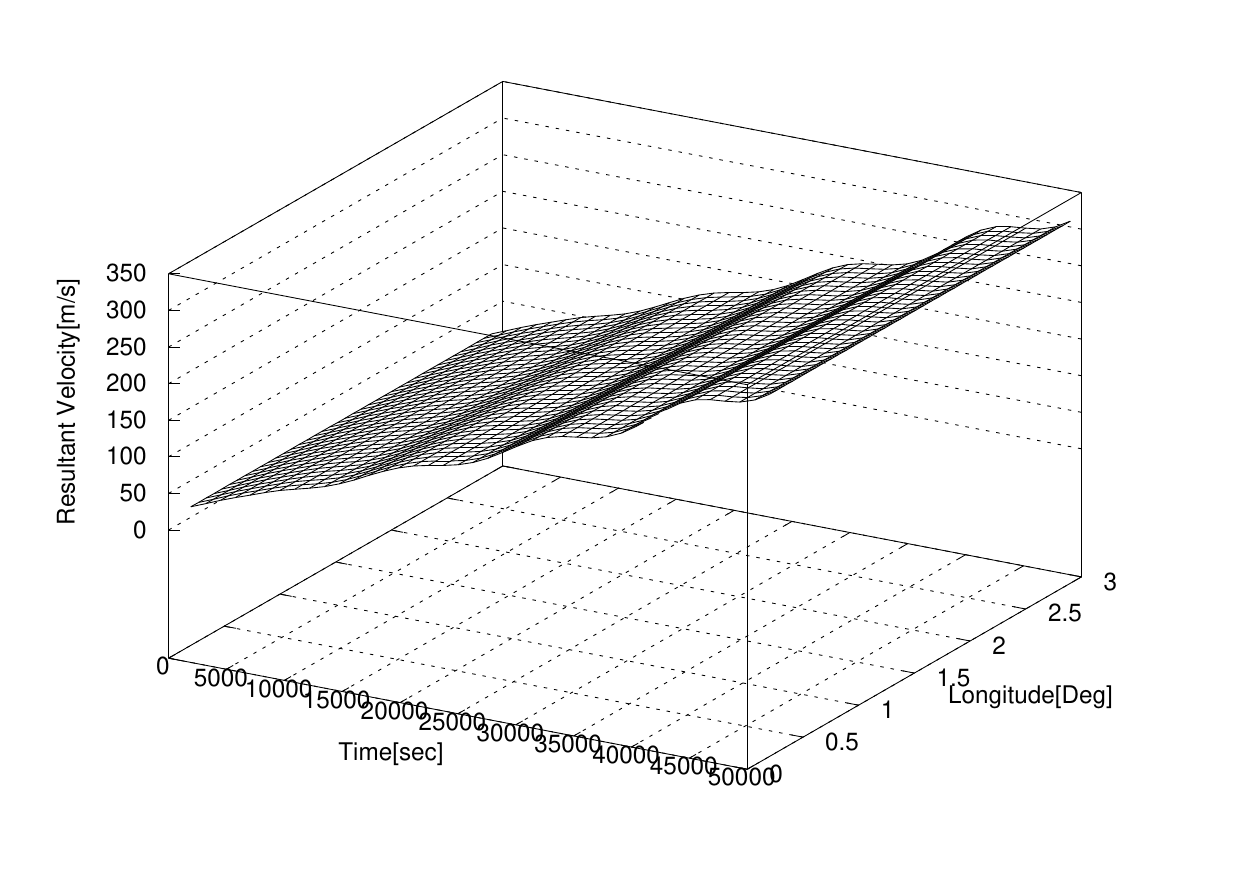}
        \caption{\label{fig3a} The plots of $\vec{v}(t,\lambda)$}
    \end{subfigure}
    \begin{subfigure}[t]{0.5\textwidth}
        \centering
        \includegraphics[height=2.40in]{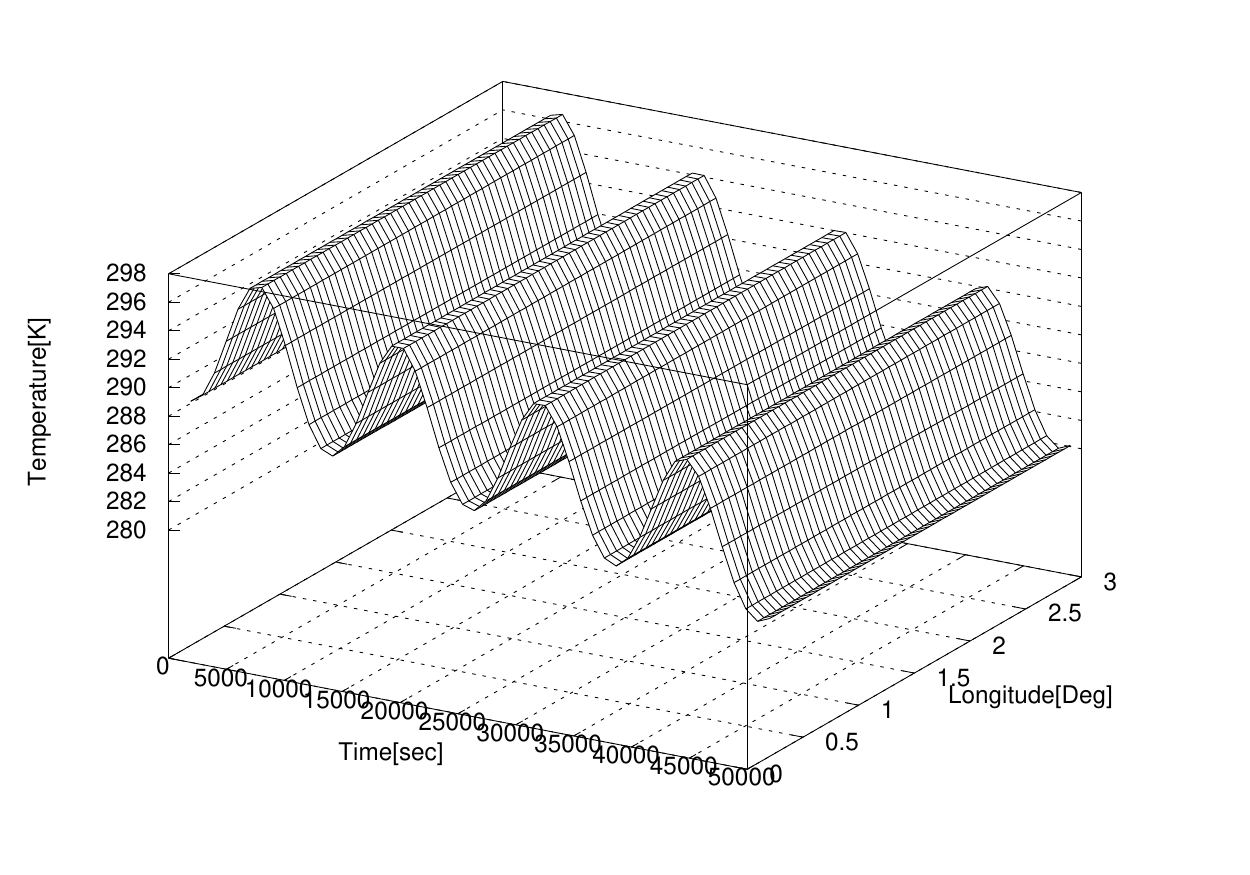}
        \caption{\label{fig3b} The plots of $T(t,\lambda)$}
    \end{subfigure}
    \caption{\label{fig3} 
    The distribution of resultant velocity(Fig.\ref{fig3a}) ,and temperature(Fig.\ref{fig3b}) with respect to time,and longitude.}
    \end{figure*}

\begin{figure*}[ht!]
     \begin{subfigure}[t]{0.5\textwidth}
        \centering
        \includegraphics[height=2.40in]{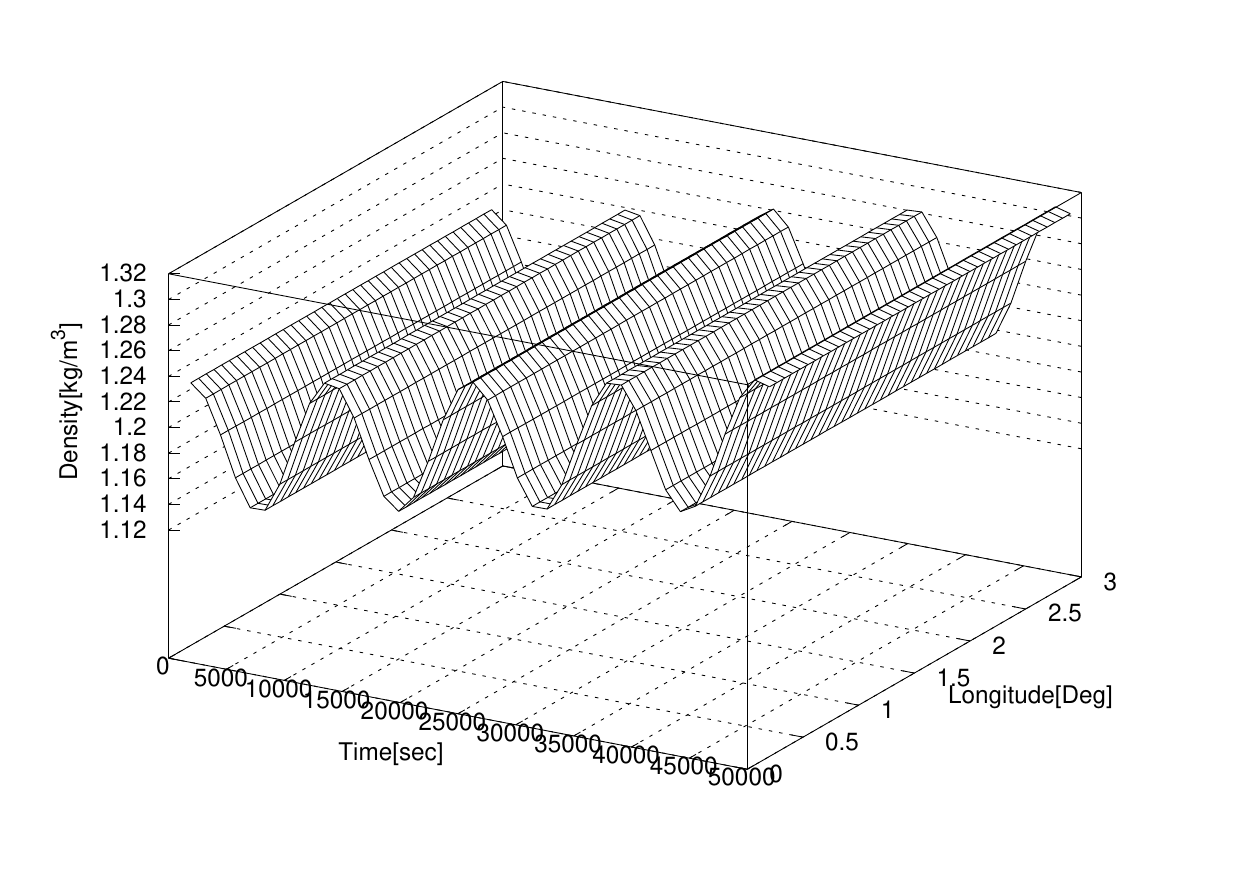}
        \caption{\label{fig4a}  The plots of $\rho(t,\lambda)$ }
    \end{subfigure}
    \begin{subfigure}[t]{0.5\textwidth}
        \centering
        \includegraphics[height=2.30in]{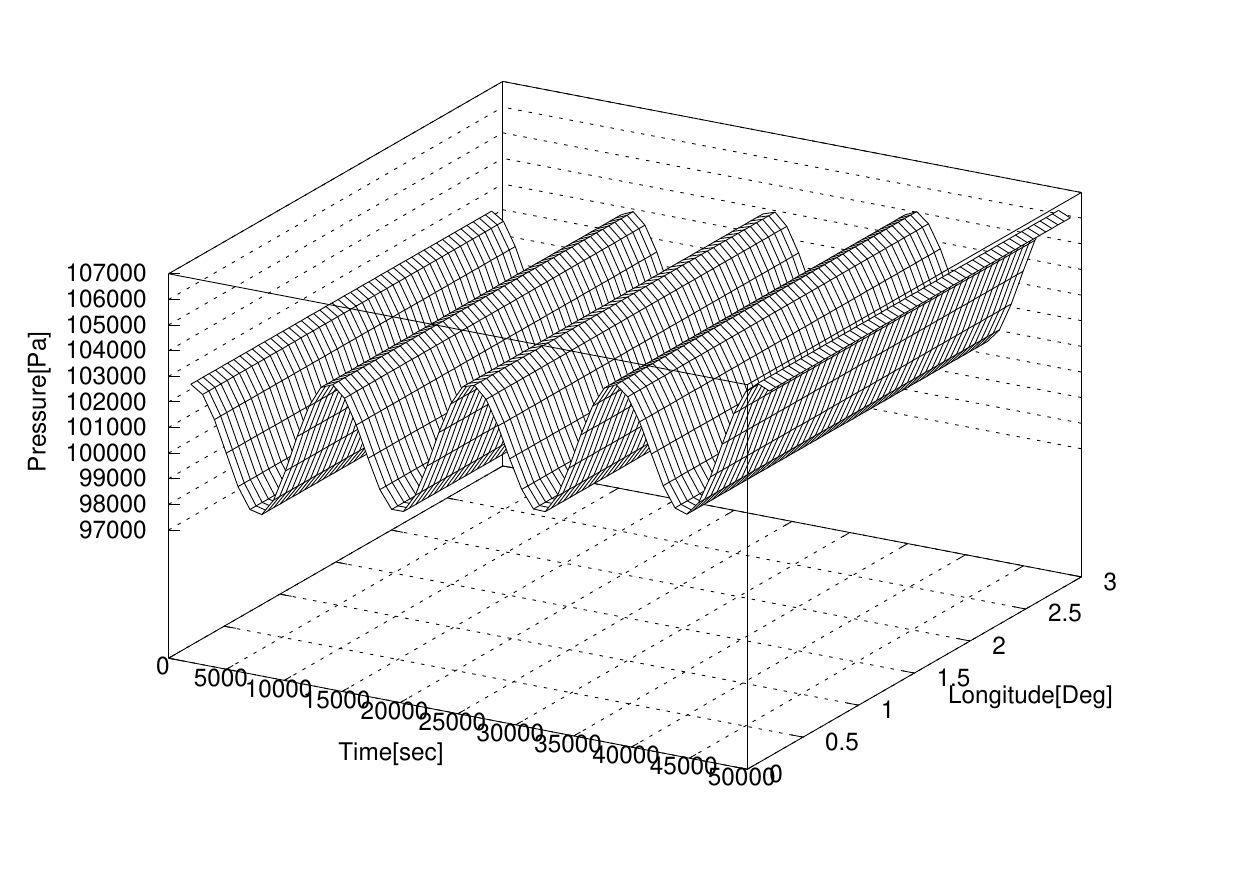}
        \caption{\label{fig4b} The plots of $P(t,\lambda)$}
    \end{subfigure}
    \caption{\label{fig4} 
    The distribution of density( Fig.\ref{fig4a}) ,and pressure( Fig.\ref{fig4b}) with respect to time,and longitude.}
\end{figure*}

\begin{figure*}[ht!]
     \begin{subfigure}[t]{0.5\textwidth}
        \centering
        \includegraphics[height=2.40in]{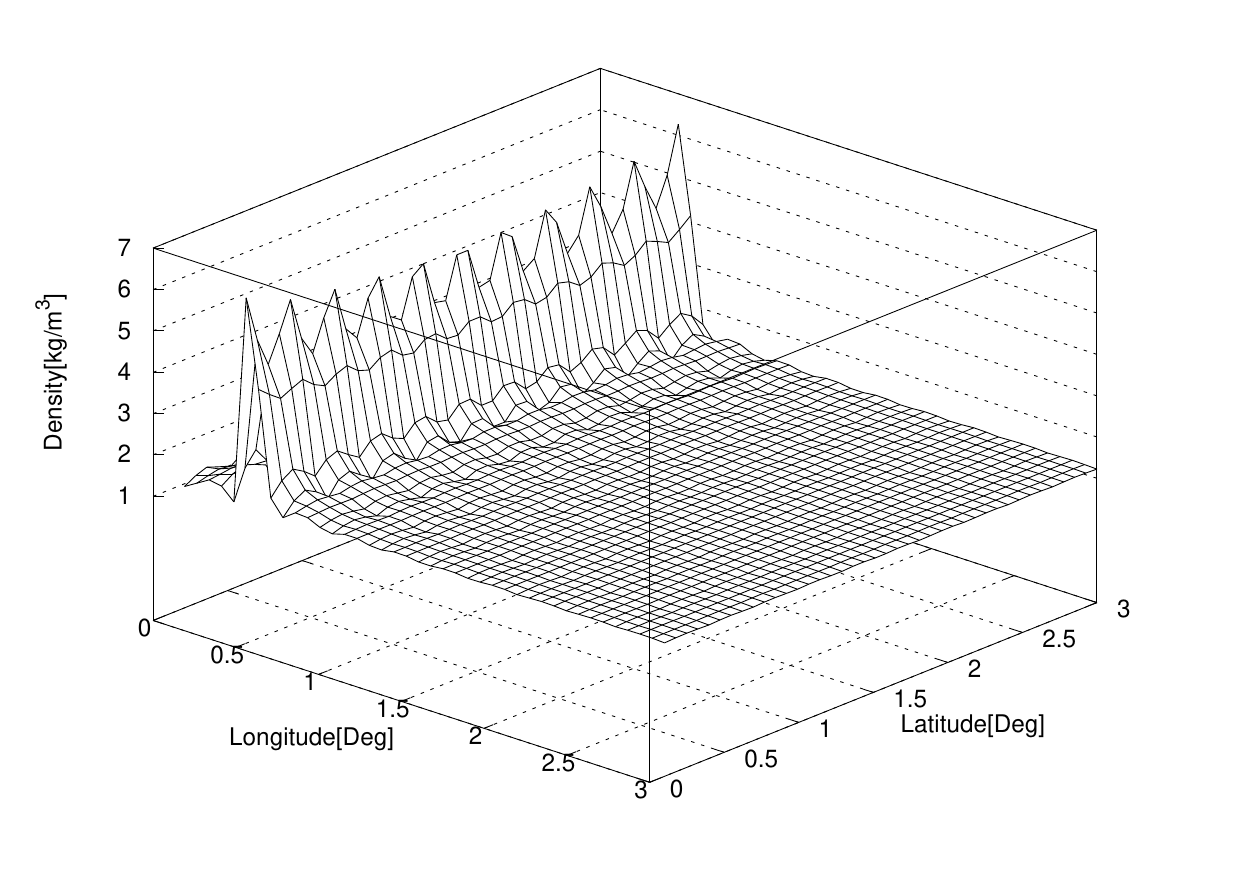}
        \caption{\label{fig5a}  The initial perturbed air mass
density at time = 0 sec at $5\Delta \lambda$. }
    \end{subfigure}
    \begin{subfigure}[t]{0.5\textwidth}
        \centering
        \includegraphics[height=2.40in]{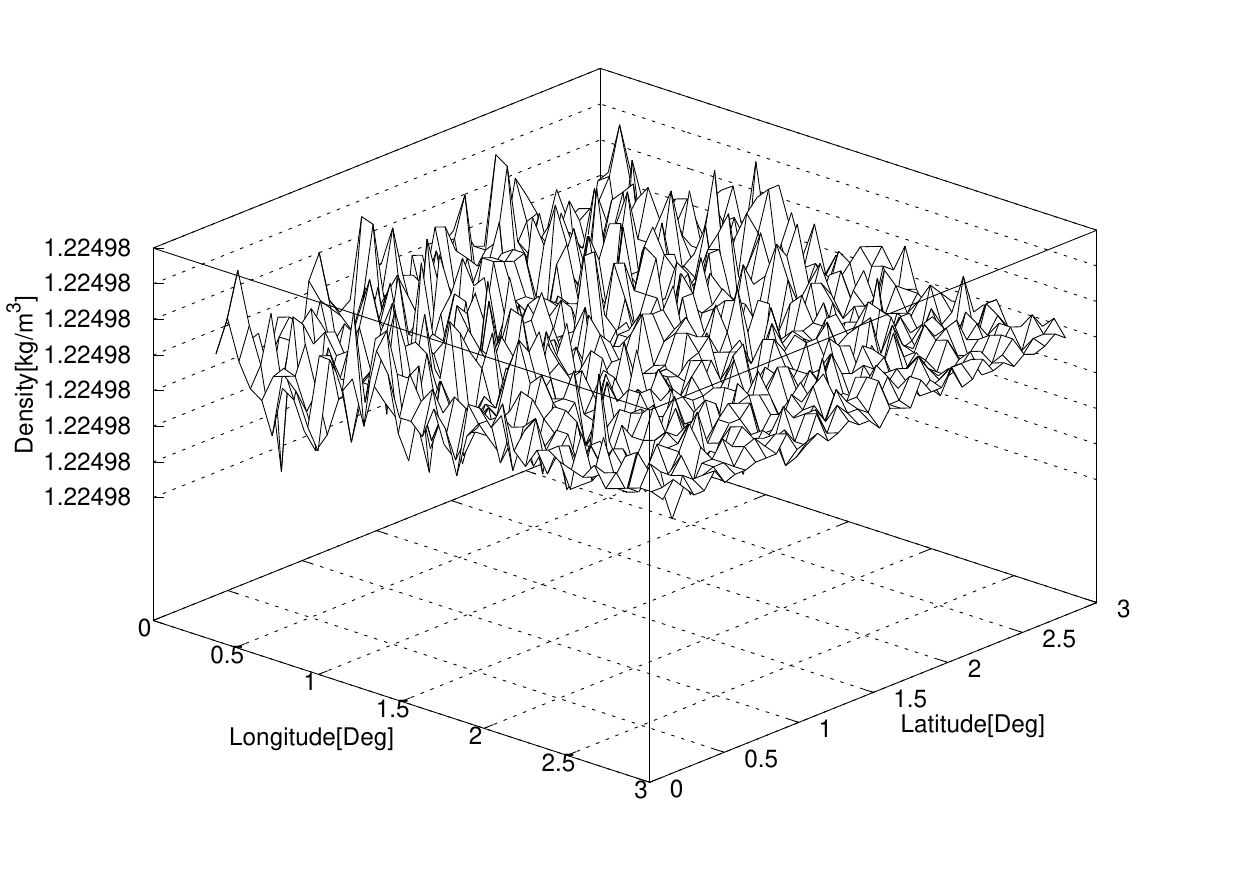}
        \caption{\label{fig5b} The  computed numerical air mass
density at time = 100 sec}
    \end{subfigure}
    \begin{subfigure}[t]{0.5\textwidth}
        \centering
        \includegraphics[height=2.30in]{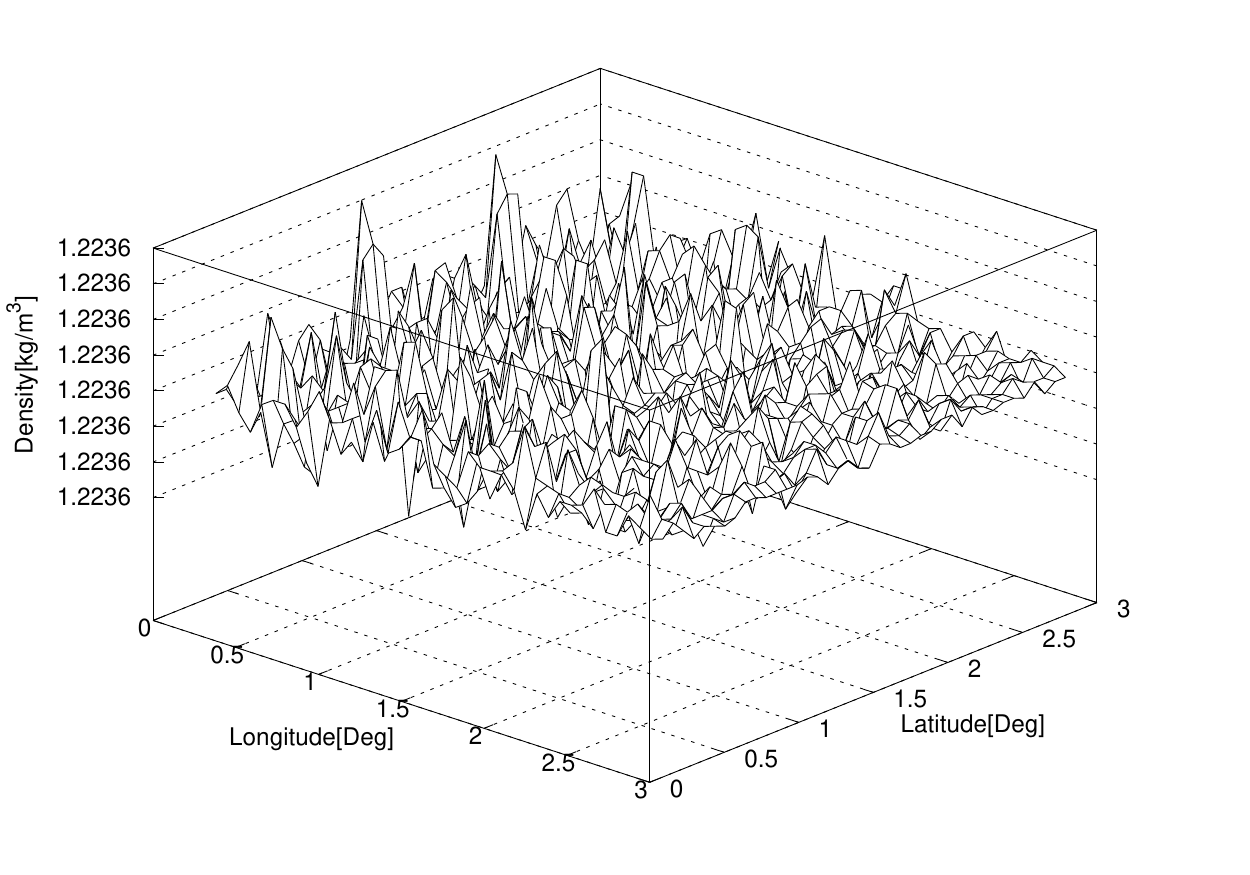}
        \caption{\label{fig5c} The  computed numerical air mass
density at time = 200 sec}
    \end{subfigure}
    \caption{\label{fig5} 
    The graph of density against longitude $\&$ latitude. From top to bottom: propagation of density at t=0sec (left column) and at t=100sec(right column), and t=200sec(left last column)}.
\end{figure*}

\begin{figure*}[ht!]
     \begin{subfigure}[t]{0.5\textwidth}
        \centering
        \includegraphics[height=2.50in]{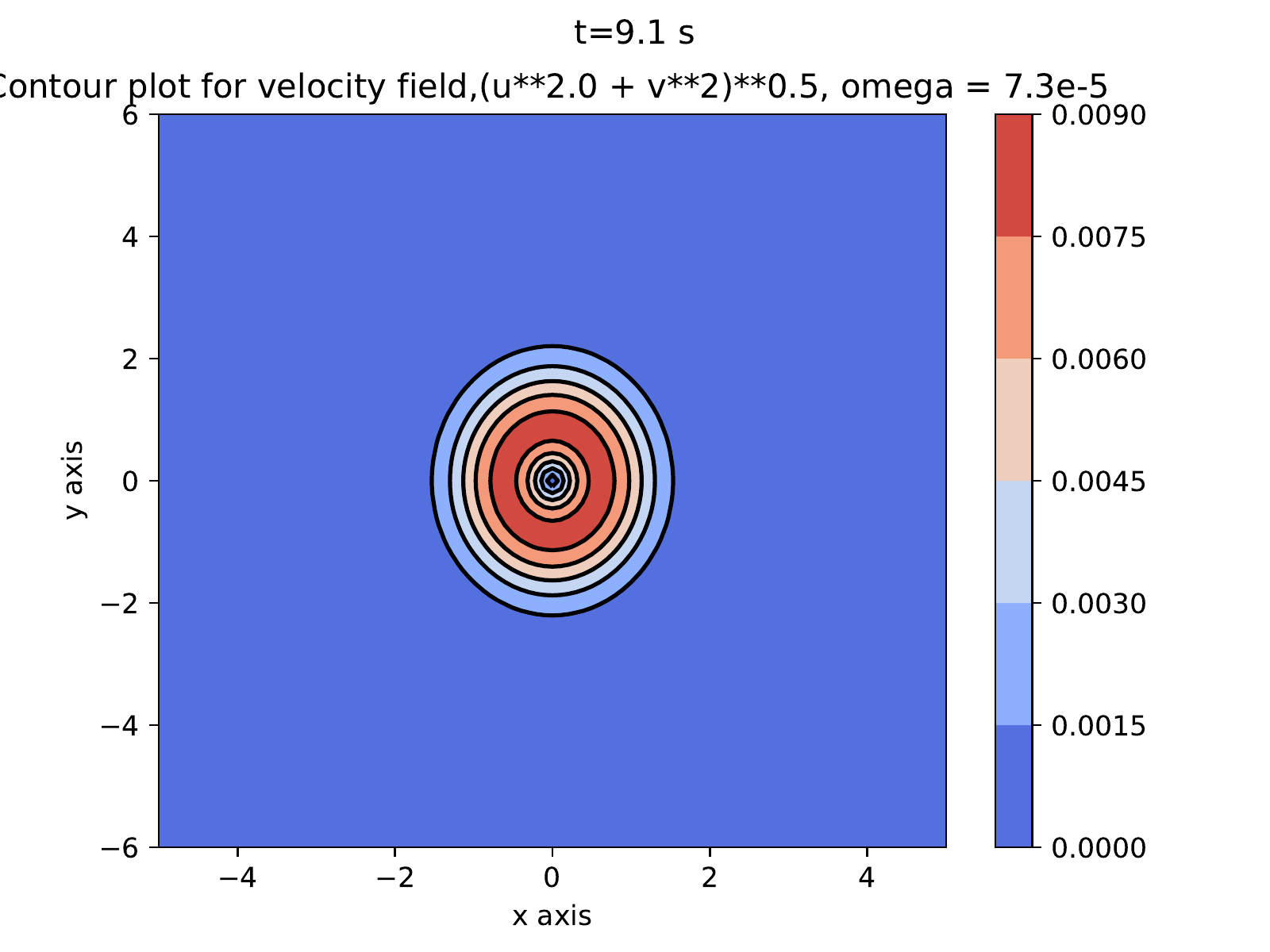}
        \caption{\label{fig6a} A contour map that describe the resultant velocity at t = 9.1sec}
    \end{subfigure}
    \begin{subfigure}[t]{0.5\textwidth}
        \centering
        \includegraphics[height=2.50in]{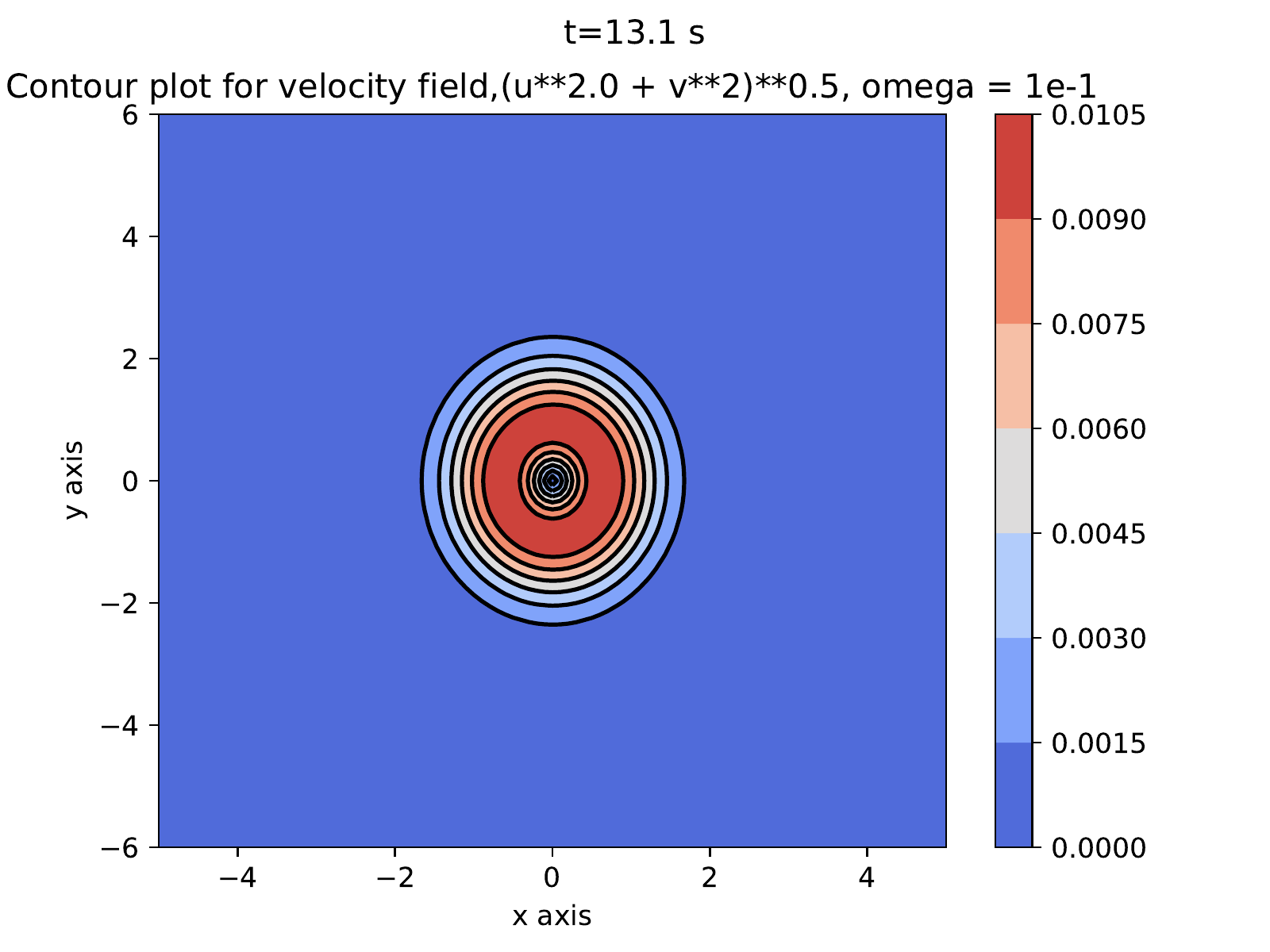}
        \caption{\label{fig6b} A contour map that describe the resultant velocity at t = 115.1sec}
    \end{subfigure}
    \begin{subfigure}[t]{0.5\textwidth}
        \centering
        \includegraphics[height=2.30in]{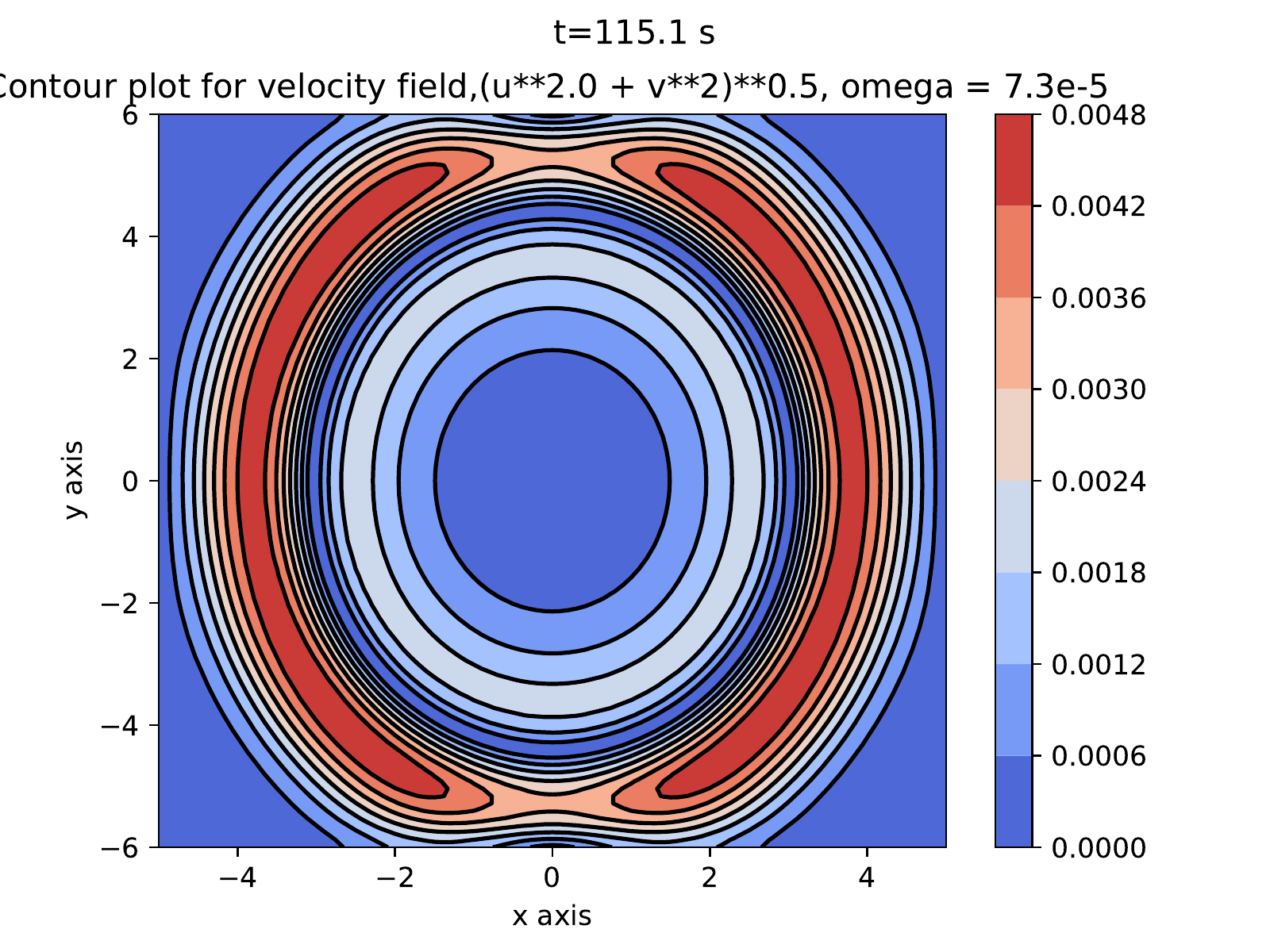}
        \caption{\label{fig6c} A contour map that describe the resultant velocity at t = 115.1sec }
    \end{subfigure}
    \begin{subfigure}[t]{0.5\textwidth}
        \centering
        \includegraphics[height=2.30in]{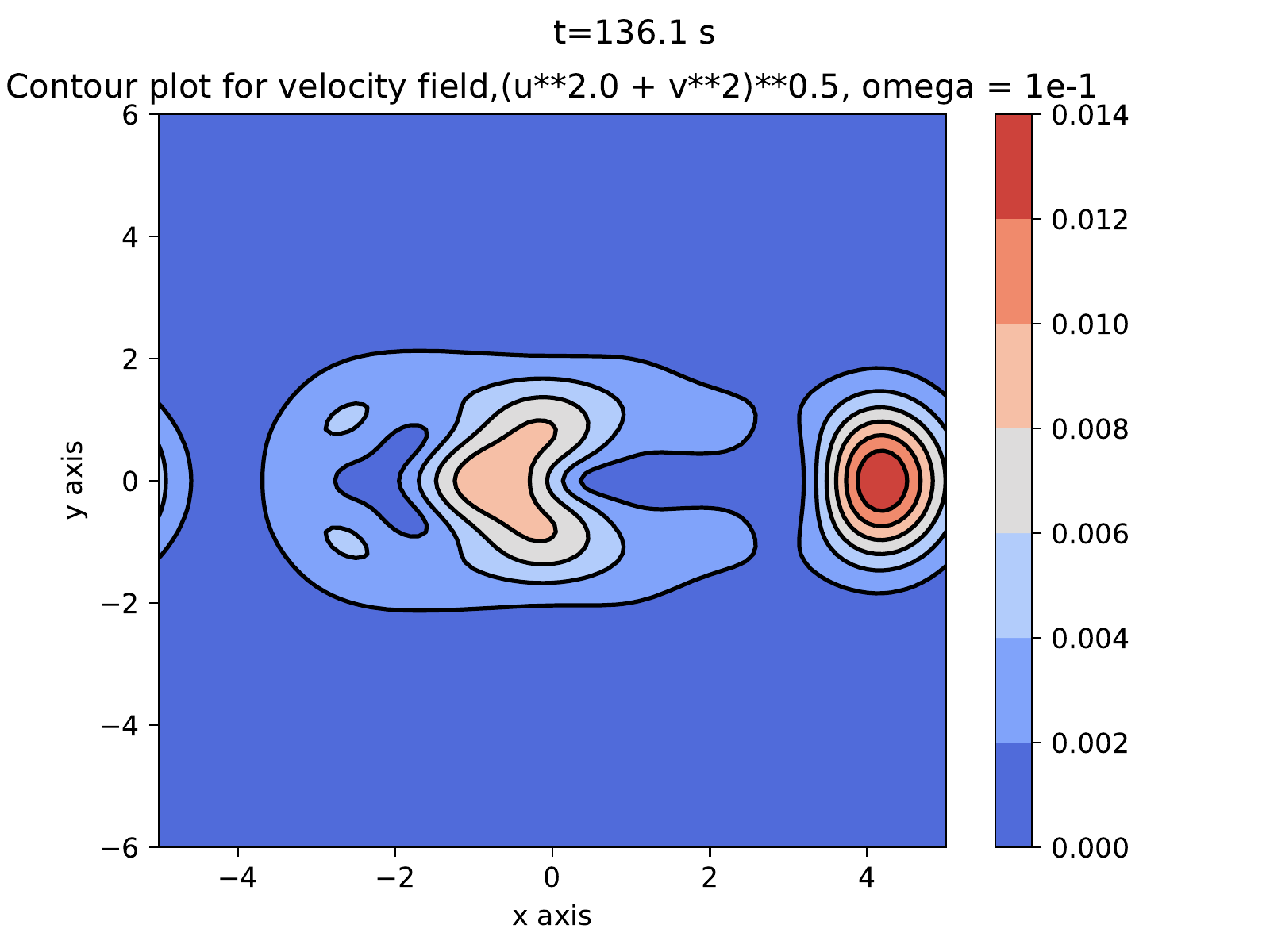}
        \caption{\label{fig6d} A contour map that describe the resultant velocity at t = 136.1sec}
    \end{subfigure}
    \caption{\label{fig6} 
    Contour plot for velocity,$\vec{U}=(u^{2}+v^{2})^{0.5}$ , From top to bottom: for small (left column) and large (right column) magnitude of Coriolis force}   
\end{figure*}

The result presented in this section
is computed thermodynamic and hydrodynamic properties
of the viscous atmospheric motion with restricting our attention to the lower part of the atmosphere as compressible neutral fluid. To see the effect of transport coefficients (viscosity and thermal conductivity) on the propagation of atmospheric resultant velocity, temperature, density, and pressure with respect to time, we perform a numerical simulation using a constant and temperature-dependent transport coefficient (viscosity and thermal conductivity). Taking these parameters into account, and based on our initial boundary condition, we obtain the propagation of resultant velocity , and temperature against time for temperature-dependent as well as for temperature-independent transport coefficient as shown in Figure\ref{fig1a}(left $1\&2$ column), and in Figure\ref{fig1b}(right $1\&2$ column), respectively. By comparing their magnitude, we obtain the change in resultant, and temperature from the corresponding two graphs , and presented in graphically as shown in the last end column of Figures\ref{fig1a}$~\&$~\ref{fig1b}. Thus, from those corresponding figures, we have observed that there is a slight change in both resultant velocity and temperature with respect to time when we used temperature-dependent transport coefficient instead of temperature-independent transport coefficient. 
\\
 Moreover, the propagation of density and pressure with respect to time is becomes 
 as shown in Figure\ref{fig2a}(left column) and \ref{fig2b}(right column) for both temperature-dependent ($\mu(T),K(T)$) and temperature-independent transport coefficient ($ ~\mu,k$). Again here, their corresponding variation with respect to time is presented in graphically as shown in Figure \ref{fig2a} (left last column) and Figure\ref{fig2b}~(right last column), respectively. Based on these results, a correction, should be always considered for the computation of thermodynamic and hydrodynamic properties of the viscous atmospheric motion, and for better accuracy, temperature-dependent thermal conductivity and coefficient of viscosity should be considered.  Hence, we consider the temperature-dependent transport coefficient for numerical solving all of the governed equations that interpret our model.  
\newpage
\noindent
Depending on the above result, that is by taking temperature-dependent transport coefficient, we obtain the numerical solution for the governed equations and it is presented in graphically as shown in below Figures~\ref{fig3}$\&$\ref{fig4}. From those figures, we have seen that all the atmospheric parameters profile behaves like a wave with respect to time, and the flow patterns of Figures are very close to those shown in\cite{bruneau20062d}. The propagation of resultant velocity and the temperature have opposite in directions as we observe, in Figure\ref{fig3}(\ref{fig3a} ,$\&$ \ref{fig3b}),and the reasons for this is currently is due to the dissipation of the fluid their is a transformation mechanical energy  to the thermal and a vice verse relation  .
    \\
    When we apply the finite difference method on continuity equation, we obtain the propagation of density($\rho$) as function of geometrically positions and the time taken. Here, the propagation of density was presented graphically as shown in Figure\ref{fig4a} with respect to time and longitude, and the propagation of atmospheric pressure  as shown in Figure\ref{fig4b}. Thus, both of the density and the pressure graph oscillate with same direction respect to time . Due to ideal gas relation,  from Figures\ref{fig3b},\ref{fig4a},$\&$ \ref{fig4b} we have seen that
    the propagation of temperature is also in opposite direction to the propagation of density and pressure of the fluid .
\newpage
\noindent 
When a sort of horizontal air mass density, $\rho\approx10kg/m^{3}$ occurred as a perturbation  on the atmosphere at a specific longitude then this perturbation of the density change its propagation with time as shown in Figure\ref{fig5} in the form of full three dimensional surface. Thus, from the first figure\ref{fig5a}, we observed that propagation of density has a maximum amplitude at perturbation point , then its oscillation becomes decay to atmospheric density at sea level $\rho=1.225kg/m^{3}$ in both direction of longitude. When the time taken reaches to 100sec and 200sec,the propagation of density behave like a random motion with increasing in amplitude as shown in figures(\ref{fig5b}, $\&$ \ref{fig5c}). In similar fashion,using this perturbed density value, we also determine the propagation for  the remaining of the atmospheric parameter  along to the geometrically position (longitude,and latitude) at specific time taken (see appendices  with  detailed). Finally, we can put observed that the propagation of resultant velocity,temperature,and pressure behave like random motion with respect to longitude and latitude at
different time moments.
\newpage
\noindent
Finally, it is interesting to note that the Coriolis effect does not
appear in the radial component of Naiver-Stokes equation which is related to two-dimensional
velocity. Its effect appears on the equations for  resultant velocity.
Hence, by contrast the magnitude of the rate of momentum change  due to Coriolis  forces that exerted on the atmospheric compressible fluid, we obtain the resultant velocity field shown as a color plot with some contour lines for corresponding equation\eqref{eq5} - \eqref{eq6} at a different time of the simulation as shown in below figures \ref{fig6} (left , and right columns) for small and large magnitude of Coriolis force.\\
\\
Figure\ref{fig6a}, and \ref{fig6b} , has been pointed out that there is an initial circular shape wave propagation was observed at the center at  around $t\approx10$ sec and due to double periodic condition, the circular wave spread out in two dimension and we mark an interesting wave phenomena with symmetric in shape observed at figure ~\ref{fig6c}, $\&$ \ref{fig6d} for respective  magnitude of Coriolis force . Hence,as the time taken increases   its magnitude to $115.1$ sec and to $136.1$ sec for corresponding magnitude of Coriolis force,  we have examined that the propagation of the wave is only along to longitude of the Earth as
illustrated in figure \ref{fig6d} for large magnitude  of the Coriolis force, and this figure \ref{fig6} general show how the resultant velocity depends on the magnitude of Coriolis force.\\
 \newpage
 \noindent
\section{Summary and Conclusions}\label{conclu}
 In this article, we propose an efficient computational strategy to deal with thermodynamic and hydrodynamic properties of the viscous atmospheric  motion in two dimension with considering temperature-dependent viscous coefficient. The dynamics of the atmosphere, governed by partial differential equation
without any approximation ,and without considering latitude-dependent acceleration due to gravity. The numerical
solution for those governed equations was solved by applying the finite difference method with applying some sort
of horizontal air mass density as a perturbation to the atmosphere at a longitude of $5\Delta\lambda$ . Based on this initial
boundary condition with taking temperature-dependent transport coefficient in to account, we obtain the propagation for
each atmospheric parameter and presented in graphically as a function of geometrically position and time. All of the parameters oscillating with respect to time and satisfy the characteristics of atmospheric wave.
\\
\\
 Finally, the effect of the Coriolis force on resultant velocity were also discussed by plotting
contour lines for the resultant velocity for different magnitude of Coriolis force, then we also obtain an interesting wave
phenomena for the respective rotation of the Coriolis force. Generally, the above research result highly sensitive to the initial given value of the parameters ,longitude and latitude grid size,and on time scale.
Our future work shall be towards evolution of three dimensional governed equation taking compressible ionized
fluid with latitudinal dependent acceleration due to gravity.
\section*{Acknowledgments}
The first author  gratefully acknowledges financial support from Addis Ababa University and Mizan-Tepi University for enabling him to carry out this research. The authors would also like to thank The International Science Programme, Uppsala University, Sweden for the support they have provided to our research group.
\bibliography{nse2.bib}
\bibliographystyle{ieeetr}
\newpage
\noindent
\section{Appendix}
\subsection{The distribution of atmospheric parameters with respect to geometrically position }
\begin{figure*}[ht!]
     \begin{subfigure}[t]{0.5\textwidth}
        \centering
        \includegraphics[height=2.40in]{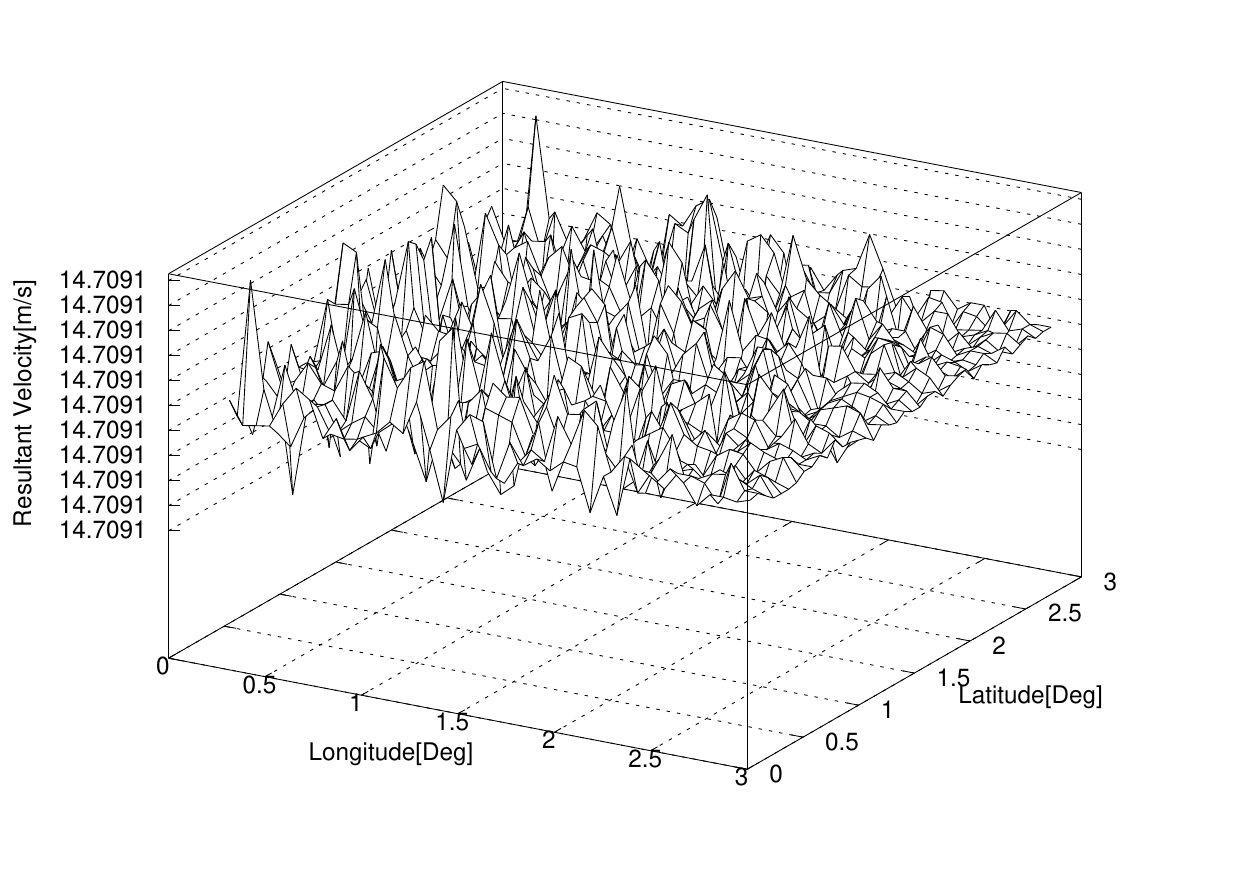}
        \caption{\label{fig7a} The plots of $\vec{v}(t,\lambda)$ at $t=100sec$}
    \end{subfigure}
    \begin{subfigure}[t]{0.5\textwidth}
        \centering
        \includegraphics[height=2.40in]{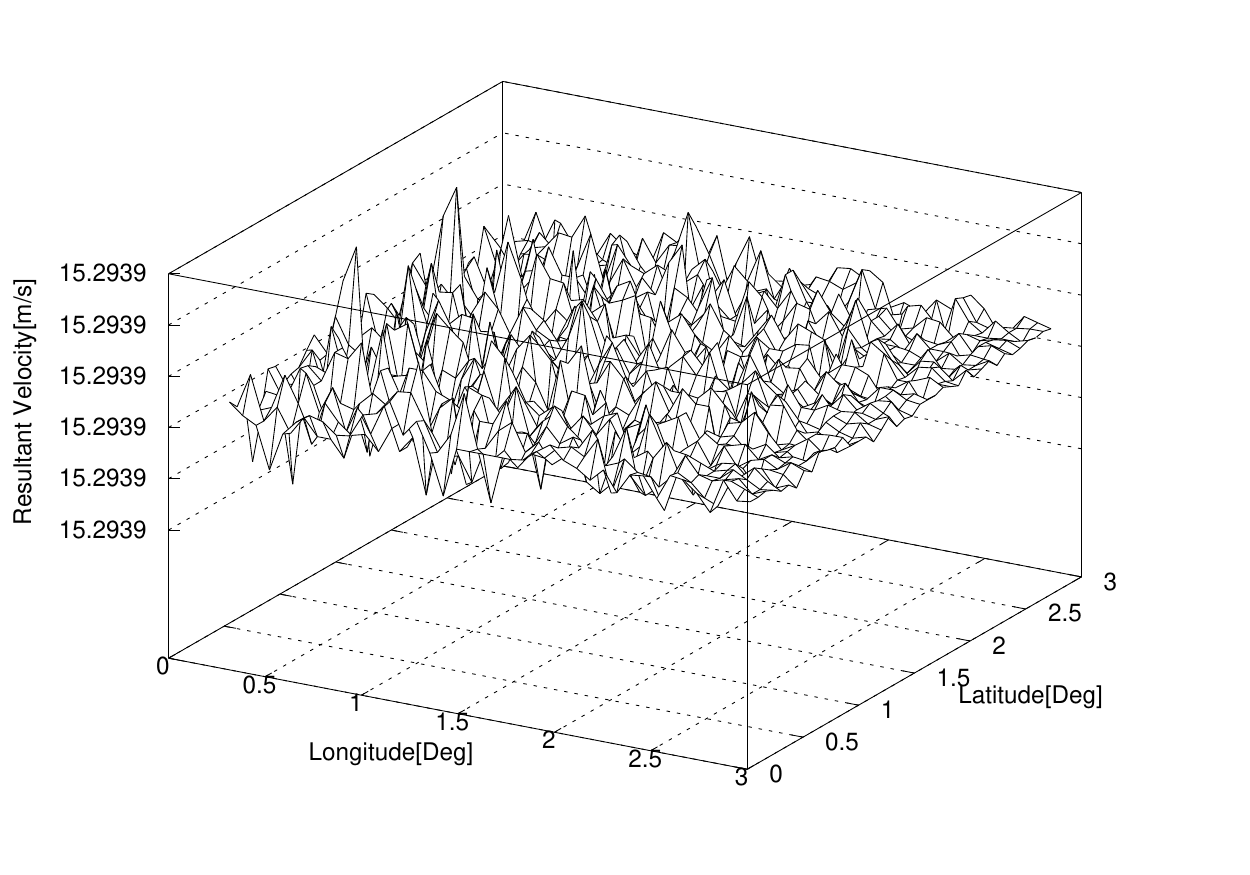}
        \caption{\label{fig7b} The plots of $\vec{v}(t,\lambda)$ at $t=200sec$}
    \end{subfigure}
    \caption{\label{fig7} 
    The distribution of resultant velocity at 100sec(Fig.\ref{fig7a}) ,and 200sec(Fig.\ref{fig7b}) with respect to longitude, and latitude.}
    \end{figure*}
    \begin{figure*}[ht!]
     \begin{subfigure}[t]{0.5\textwidth}
        \centering
        \includegraphics[height=2.40in]{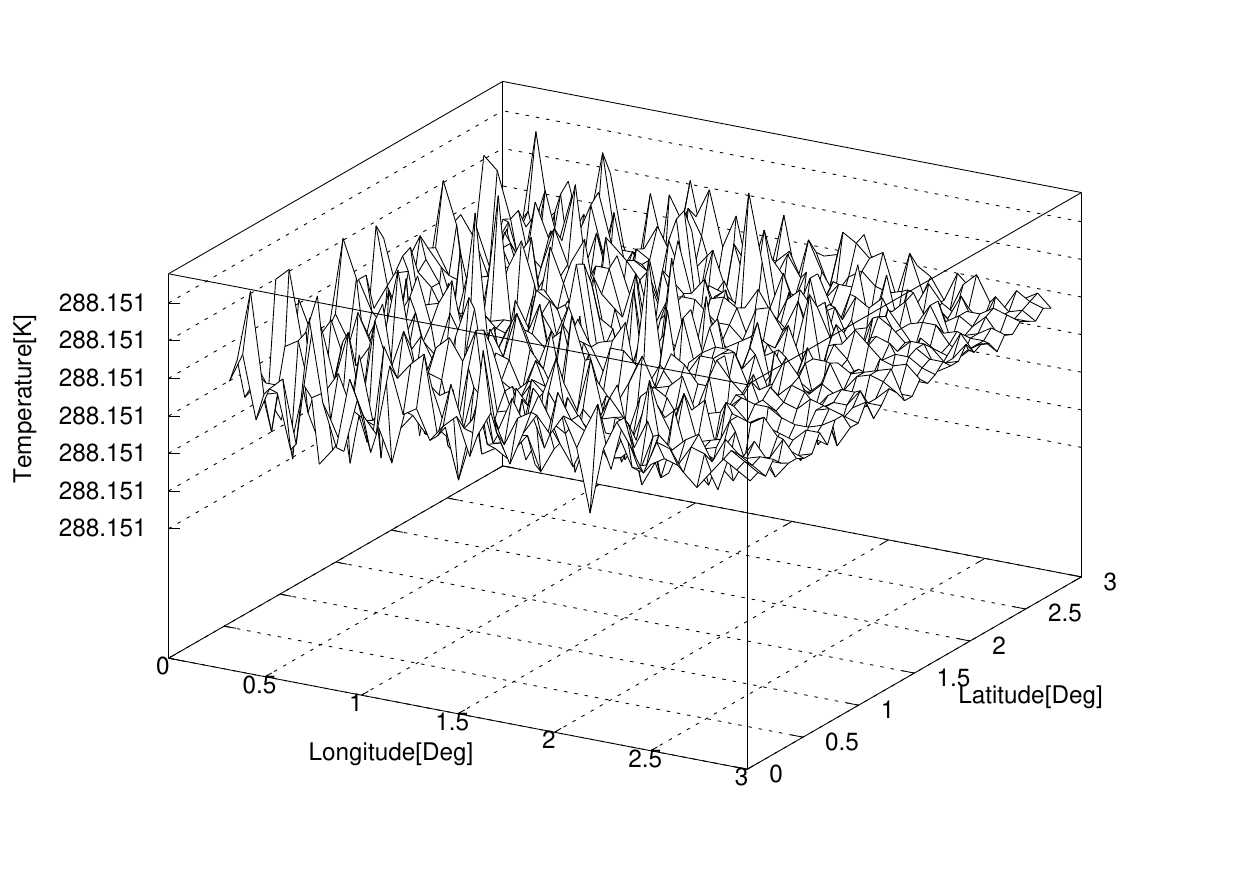}
        \caption{\label{fig8a} The plots of $T(t,\lambda)$ at $t=100sec$}
    \end{subfigure}
    \begin{subfigure}[t]{0.5\textwidth}
        \centering
        \includegraphics[height=2.40in]{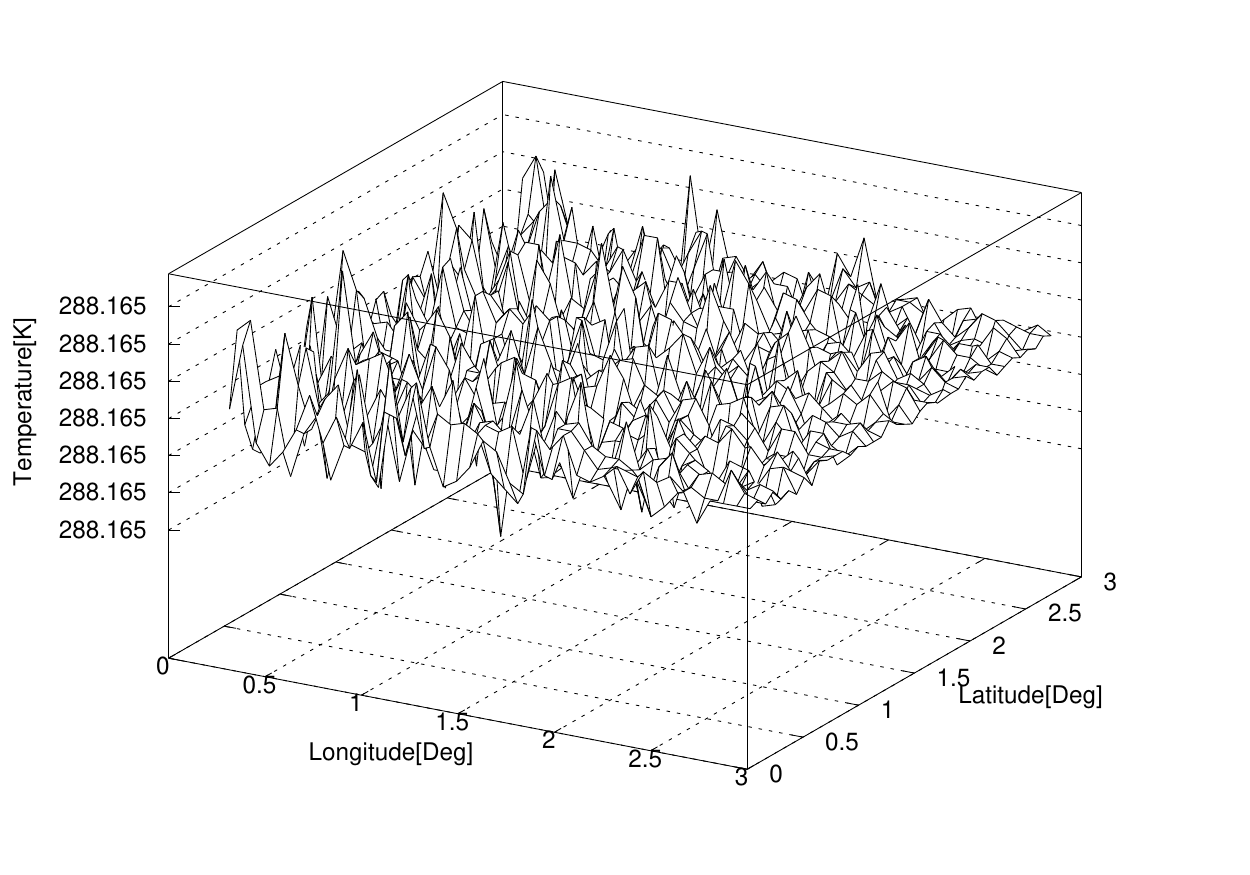}
        \caption{\label{fig8b} The plots of $T(t,\lambda)$ at $t=200sec$}
    \end{subfigure}
    \caption{\label{fig8} 
    The distribution of temperature at 100sec(Fig.\ref{fig8a}) ,and 200sec(Fig.\ref{fig8b}) with respect to longitude, and latitude.}
    \end{figure*}
\begin{figure*}[ht!]
     \begin{subfigure}[t]{0.5\textwidth}
        \centering
        \includegraphics[height=2.40in]{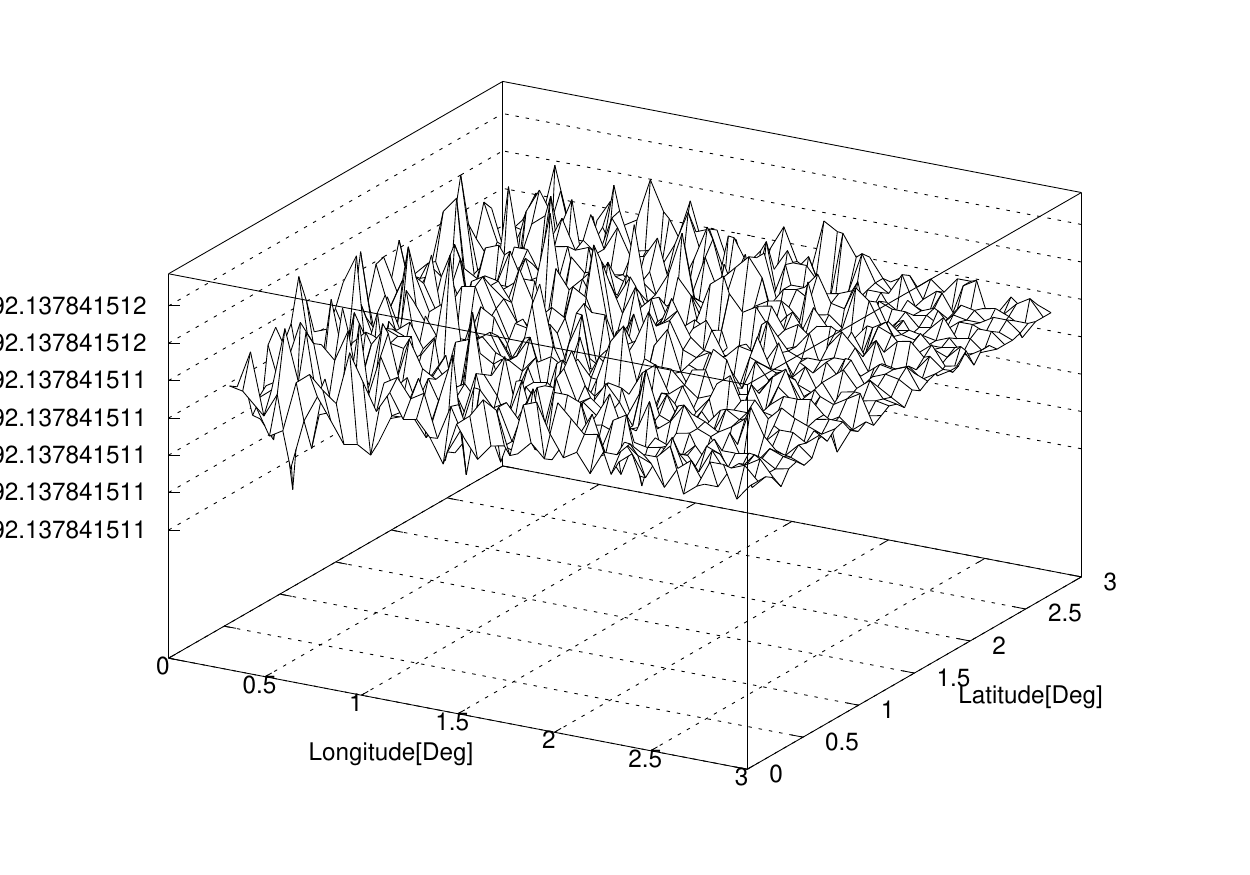}
        \caption{\label{fig9a} The plots of $P(t,\lambda)$ at $t=100sec$}
    \end{subfigure}
    \begin{subfigure}[t]{0.5\textwidth}
        \centering
        \includegraphics[height=2.40in]{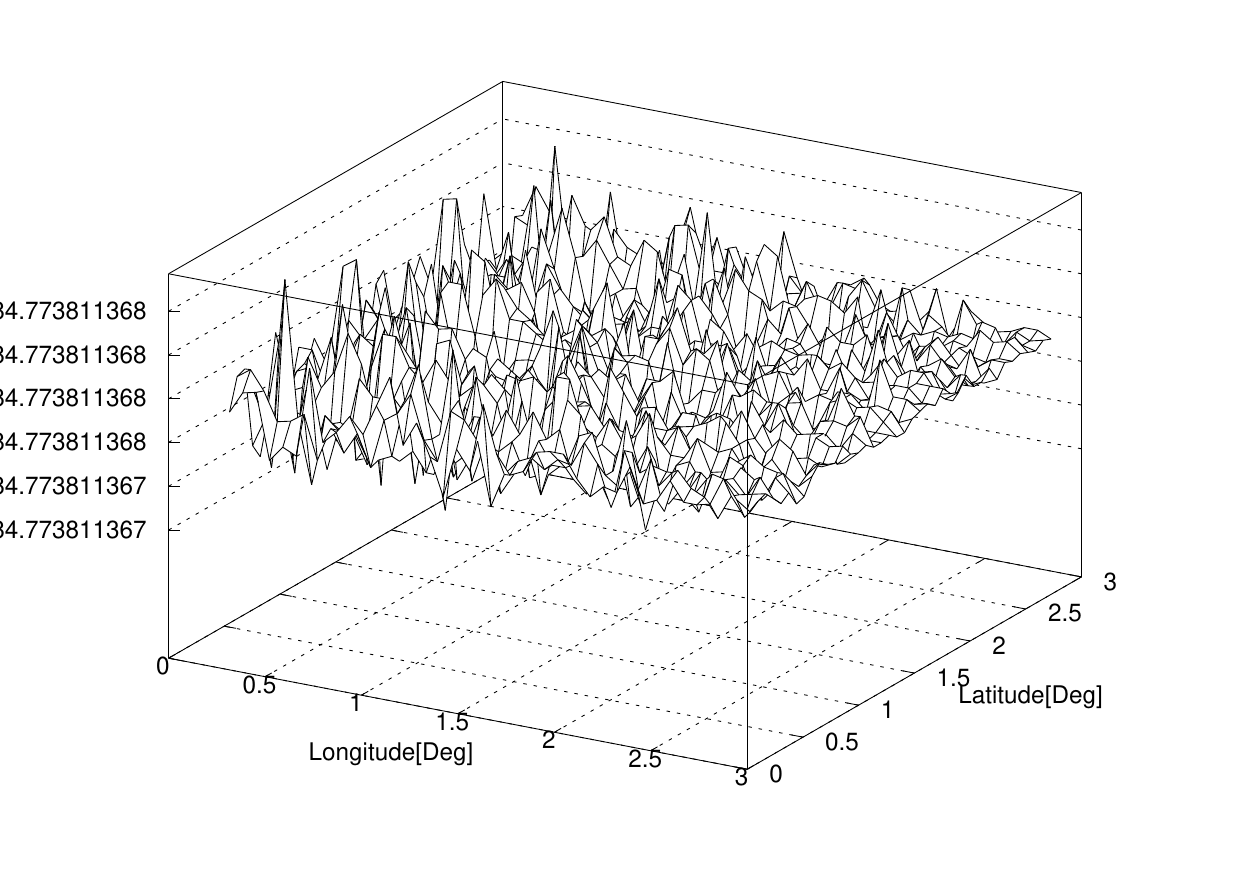}
        \caption{\label{fig9b} The plots of $P(t,\lambda)$ at $t=200sec$}
    \end{subfigure}
    \caption{\label{fig9} 
    The distribution of pressure at 100sec(Fig.\ref{fig9a}) ,and  200sec(Fig.\ref{fig9b}) with respect to longitude, and latitude.}
    \end{figure*}
  \end{document}